\documentclass[%
reprint,
aip,
showkeys,
amsmath,amssymb,
longbibliography,
]{revtex4-1}

\usepackage{graphicx}
\usepackage[caption=false]{subfig} 
\usepackage{pifont}
\usepackage{accents} 

\usepackage{dcolumn}
\usepackage{bm}

\usepackage[unicode,colorlinks=true,linkcolor=blue,citecolor=blue,urlcolor=blue]{hyperref}
\usepackage{lineno}

\newcommand{\ee}{\end{equation}} 
\newcommand{\be}{\begin{equation}}

\usepackage[mathscr]{euscript}  
\usepackage{xcolor}  

\usepackage{tcolorbox}
\usepackage{comment}
\usepackage{float}

\makeatletter
\newsavebox{\@brx}
\newcommand{\llangle}[1][]{\savebox{\@brx}{\(\m@th{#1\langle}\)}%
	\mathopen{\copy\@brx\kern-0.5\wd\@brx\usebox{\@brx}}}
\newcommand{\rrangle}[1][]{\savebox{\@brx}{\(\m@th{#1\rangle}\)}%
	\mathclose{\copy\@brx\kern-0.5\wd\@brx\usebox{\@brx}}}
\makeatother

\makeatletter
\newcommand{\vast}{\bBigg@{3.4}}
\makeatother

\usepackage{tikz}

\makeatletter
\newcommand\emailx[1]{%
	\move@AF%
	\def\@affil{{\normalfont\,#1\strut}{}}%
}%
\makeatletter
\let\auto@bib\@empty
\let\auto@bib@innerbib\@empty
\makeatother
\begin{document}

	\preprint{ApS/123-QED}

	\title{Orbit topology of clusters in active matter}

	\author{Iman Abdoli}
	\thanks{Corresponding author: Iman.Abdoli@uni-duesseldorf.de}
	\affiliation{Institut für Theoretische Physik II - Weiche Materie, Heinrich-Heine-Universität Düsseldorf, Universitätsstraße 1, D-40225 Düsseldorf, Germany}
	
	\author{Marco Musacchio}
	\affiliation{Institut für Theoretische Physik II - Weiche Materie, Heinrich-Heine-Universität Düsseldorf, Universitätsstraße 1, D-40225 Düsseldorf, Germany}
	
	\author{Lorenzo Caprini}
	\affiliation{Physics Department, University of Rome La Sapienza, P.le Aldo Moro 5, IT-00185 Rome, Italy}

	\author{Hartmut L\"{o}wen}
	\affiliation{Institut für Theoretische Physik II - Weiche Materie, Heinrich-Heine-Universität Düsseldorf, Universitätsstraße 1, D-40225 Düsseldorf, Germany}

	\begin{abstract} 
				\vspace*{10pt}
		\section*{Abstract}	
				\vspace*{-10pt}

Across living and synthetic active matter, energy-consuming units spontaneously organise into moving mesoscopic assemblies, commonly known as active clusters. Found in systems ranging from colloidal living crystals and migrating cell clusters to bacterial populations and robotic swarms, these assemblies sustain internal motion while following coherent collective trajectories. Here we introduce an orbit topology that classifies their collective motion, extending the use of topology in active matter beyond defects of ordered fields and topological excitations. We show that interacting active particles can spontaneously select topologically distinct closed trajectories, forming a discrete hierarchy of single- and multi-loop orbits characterised by an integer winding-based invariant. Self-aligning active particles under smooth anisotropic confinement provide a minimal realization, where activity couples mechanical deflections to propulsion, selecting distinct orbital sectors. Remarkably, the same invariant emerges in probability space: the centre-of-mass probability current develops one circulation cell per orbital loop, while real-space vorticity and enstrophy reveal a distinct internal organisation. Our results establish the self-sustained trajectory of a finite active assembly as a topological object and introduce a general framework for classifying active clusters through the topology of their collective motion.

	\end{abstract}


	\maketitle

	\section*{Introduction}

\begin{figure*}
	\centering
	\includegraphics[width=1\textwidth]{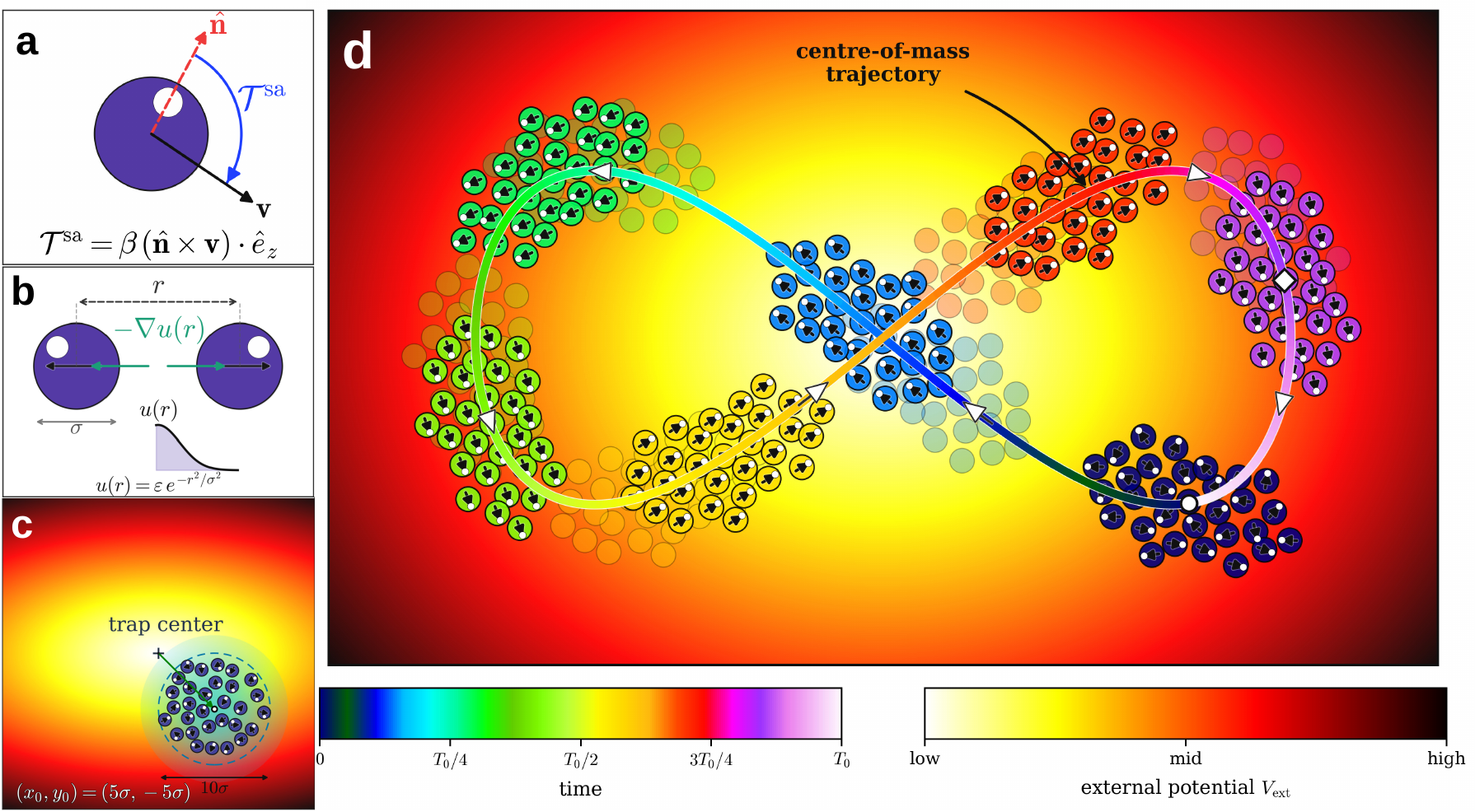}
	\caption{ \textbf{Model ingredients and collective orbital response.} \textbf{a}, Self-alignment mechanism. The propulsion direction $\hat{\mathbf n}$ rotates towards the instantaneous particle velocity $\mathbf v$ through the torque $\mathcal T^{\rm sa}=\beta(\hat{\mathbf n}\times\mathbf v)\!\cdot\!\hat{\mathbf e}_z$ , thereby coupling orientational and translational dynamics. Here $\beta$ is the self-alignment strength. The black arrow denotes the instantaneous particle velocity and the white marker the propulsion direction. \textbf{b}, Pair interaction. Particles repel through the soft Gaussian potential $u(r)=\epsilon\exp(-r^2/\sigma^2)$, with interaction strength $\epsilon$ and characteristic length $\sigma$. \textbf{c}, Initial condition. The active fluid is initialized as a localized off-centre Gaussian density distribution, explicitly breaking the left--right symmetry of the confining trap. \textbf{d}, Collective response in the anisotropic harmonic potential $V_{\rm ext}(x,y)=k_x(x^2+\alpha y^2)/2$, with $\alpha=k_y/k_x$. The background colour denotes the external potential, from low values near the trap centre to larger values away from it. Starting from the localized disordered state, the combined action of self-propulsion, repulsion, confinement and self-alignment produces a coherent many-body orbital motion. Illustrative particle configurations are shown at successive phases of one collective period $T_0$, with their colours matched to the time scale below. The corresponding centre-of-mass trajectory $\mathbf R_{\rm CM}(t)$ is shown by the time-coloured curve, with arrows indicating its direction of traversal. } 
	\label{figure01}
\end{figure*}


Spontaneous organisation across scales is a hallmark of active matter, where individually driven units collectively generate persistent structures and motion unavailable to equilibrium systems \cite{Ramaswamy2010,Marchetti2013,Shankar2022, kreienkamp2022clustering}.
A recurrent form of this organisation is the emergence of finite, spatially localized assemblies that move and reorganise as coherent mesoscopic entities while retaining non-trivial internal dynamics~\cite{KreienkampKlapp2024}. Light-activated colloids reversibly form ``living crystals'' \cite{Palacci2013}; motor-driven filaments self-organise into coherently moving structures and vortex lattices \cite{Sanchez2012,Sumino2012}. Motile colloids likewise develop macroscopic directed motion and collective vortices \cite{Bricard2013,Bricard2015}, while cohesive active condensates can migrate, fragment and reorganise under heterogeneous activity fields~\cite{vahid2026self}. Beyond these colloidal and reconstituted active systems, cohesive cell assemblies display collective migration and geometry-dependent motion \cite{Trepat2018,Vercruysse2024}, multicellular spheroids sustain coherent orbital dynamics \cite{Kim2026Spheroids}, and confined bacterial populations organise into persistent vortical states \cite{Wioland2013,Liu2021}. The same organising principles also emerge in robotic swarms, where autonomous collective motion arises through local mechanical interactions and feedback without centralized control \cite{Arbel2025Transport,Fersula2026Phototaxis}. We use the term \emph{active cluster} for this broad class of finite, spatially localized interacting assemblies that possess a well-defined collective coordinate while retaining internal active degrees of freedom, making the assembly itself an autonomous dynamical object rather than one tied to a particular microscopic realization or binding mechanism.

The persistent trajectories of finite active assemblies introduce a topological structure that is largely absent from the current description of active matter. Topology has so far been associated primarily with defects of ordered fields---including vortices, asters, disclinations and defect loops---whose winding reflects the structure of polar or nematic order \cite{Shankar2022,Mermin1979,Keber2014,DeCamp2015,Doostmohammadi2018,Shankar2018,ShankarMarchetti2019}, and with topological excitations and dynamical spectra, such as protected sound modes, Chern-type edge states and non-Hermitian phenomena \cite{Souslov2017,Shankar2017,Souslov2019,Sone2019,Fruchart2021}. In active systems, these structures acquire intrinsically nonequilibrium dynamics, becoming motile, proliferating, ordering collectively and even organising biological processes in tissues and developing organisms \cite{Saw2017,Kawaguchi2017,MaroudasSacks2021,Duclos2020}. A different topological object emerges when the invariant is carried not by an internal field or excitation spectrum, but by the collective trajectory of the active assembly itself. Winding statistics have previously been used to characterise stochastic paths around prescribed vortical structures \cite{Mahault2022}, while, at the single-particle level, self-alignment under anisotropic confinement is known to generate multiple orbital states and noise-driven transitions between them \cite{Damascena2022,Damascena2023}. Here, the interacting cluster generates and sustains the many-body trajectory whose connected winding domains define the orbital sector. Such recurrent states have no generic counterpart in overdamped passive dynamics, where conservative forces drive relaxation towards equilibrium and detailed balance precludes sustained probability circulation. Continuous activity removes this constraint and permits persistent currents and time-dependent collective states \cite{Ramaswamy2010,Marchetti2013,Doostmohammadi2016, kreienkamp2025synchronization}, placing the selection of orbital sectors under the control of how active dynamics couples to the surrounding geometry.

The selection of collective dynamical states is strongly shaped by the environment in which active matter evolves. Confinement can stabilize bacterial vortices and macroscopic rotation in motile colloids \cite{Wioland2013,Bricard2015}, reorganise active-nematic flows and cellular shear states \cite{Doostmohammadi2016,Duclos2018}, and, when patterned or dynamically controlled, localize stresses and direct active transport \cite{Peng2016,Ross2019}. Geometry also acts on living active assemblies: curvature generates propagating velocity waves in rotating tissues \cite{Brandstatter2023SphericalTissues}, the shape of epithelial clusters controls their autonomous migration \cite{Vercruysse2024}, and multicellular spheroids sustain coherent orbital motion before invasion \cite{Kim2026Spheroids}; even the rheology of the surrounding medium can reorganise bacterial motion from turbulent to ordered states \cite{Liu2021}. These observations show that geometry does not merely constrain active motion, but can actively select how it is organised. A particularly direct route to such selection is provided by \emph{self-alignment}, where the propulsion polarity responds to the particle's own velocity rather than to neighbouring orientations \cite{Baconnier2025}. Mechanical forces can then feed back onto propulsion by deflecting the velocity and thereby reorienting the polarity. Across scales, this simple feedback supports confined orbital and clustered states \cite{Dauchot2019,Damascena2022,Damascena2023,Canavello2024}, reorganises dense active phases and crystals \cite{Musacchio2025MIPS,Musacchio2026Flocking,Musacchio2026Circling}, and drives collective actuation and autonomous transport in active solids and robotic swarms \cite{Baconnier2022ActiveSolids,Arbel2025Transport,Fersula2026Phototaxis}. Self-alignment therefore provides a minimal physical mechanism through which environmental geometry can be converted into autonomous collective motion. 


Here we show that local active feedback can do more than select a mode of motion: it can select the \emph{topology} of the collective trajectory itself. In a minimal realization of interacting self-aligning active particles under smooth anisotropic confinement, the same finite cluster displays a remarkably rich dynamical landscape, spanning arrested, vortex-like, stable orbiting, and oscillatory states. Their selection is strongly non-monotonic and exhibits re-entrant behaviour as the alignment strength and confinement anisotropy are varied. Within the stable-loop regime, an additional discrete organisation emerges in the form of topologically distinct single- and multi-loop trajectories. We quantify this structure by an integer orbital loop number, $\mathcal{Q}_{\rm loop}$, constructed from the winding domains of the centre-of-mass path. The same integer emerges in probability space, where the centre-of-mass probability current forms one circulation cell per orbital loop, whereas the real-space current develops a distinct internal vortex organisation.

We establish these results from particle-resolved Brownian dynamics of the microscopic model and formulate, to our knowledge, the first dynamical density-functional theory (DDFT) for self-aligning active particles. By retaining the feedback between translational currents and propulsion orientation, the theory captures the dynamical-state landscape, orbital hierarchy and internal-flow organisation while providing a general continuum framework for velocity--polarity feedback in interacting active matter. These results establish the self-sustained trajectory of a finite active assembly as a topological object and orbit topology as a general organising principle for autonomous cluster motion. Defined solely by the collective trajectory, this classification transcends the microscopic origin of the activity and points to a universal description of recurrent motion in finite active assemblies.

\section*{Results}
\begin{figure*}[t]
\centering
\includegraphics[width=1.0\textwidth]{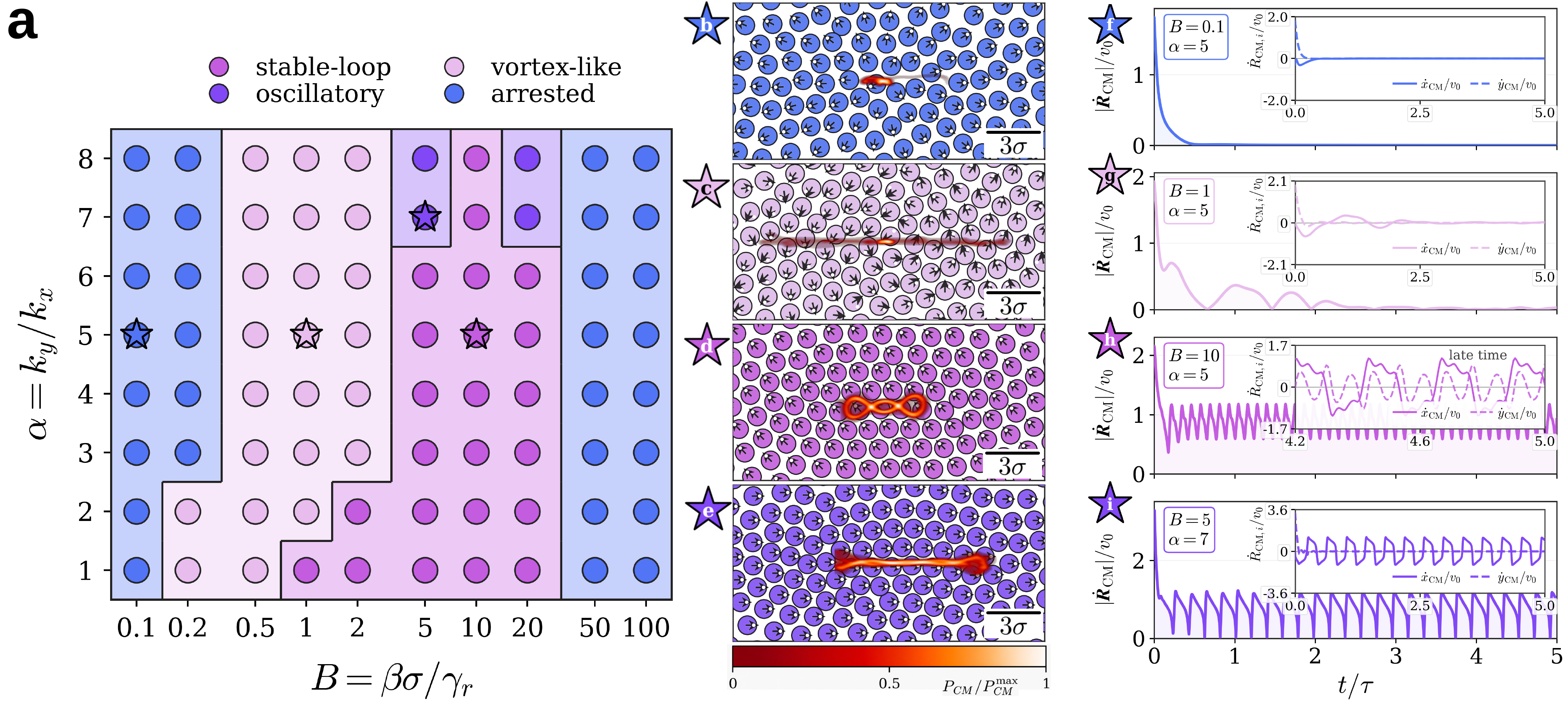}
\caption{\textbf{Dynamical state diagram and characteristic collective responses of self-aligning active Brownian particles.} \textbf{a}, Dynamical state diagram obtained from Brownian-dynamics (BD) simulations in the plane of self-alignment strength $B=\beta\sigma/\gamma_r$ and trap anisotropy $\alpha=k_y/k_x$. Colours distinguish arrested, vortex-like, stable-loop, and oscillatory states; symbols mark sampled parameters and shaded regions guide the dynamical regimes. Stars indicate the four representative states shown in \textbf{b--i}. \textbf{b--e}, Zoomed-in particle-resolved configurations and normalised CM probability density $P_{\rm CM}/P_{\rm CM}^{\max}$ from BD simulations. The arrested state (\textbf{b}) shows predominantly radially outward orientations with a localized CM, the vortex-like state (\textbf{c}) internal circulation with weak CM displacement, the stable-loop state (\textbf{d}) coherent orbital motion along a closed CM path, and the oscillatory state (\textbf{e}) shows strongly directed collective motion collectively oscillating in $x$ direction. Black arrows denote particle velocities and white markers propulsion directions; scale bars correspond to $3\sigma$. \textbf{f--i}, Corresponding dimensionless CM speed $|\dot{\mathbf R}_{\rm CM}|/v_0$ measured from DDFT, for \textbf{f} $B=0.1$, $\alpha=5$, \textbf{g} $B=1$, $\alpha=5$, \textbf{h} $B=10$, $\alpha=5$, and \textbf{i} $B=5$, $\alpha=7$. The insets show $\dot{x}_{\rm CM}/v_0$ and $\dot{y}_{\rm CM}/v_0$ and distinguish different dynamical states: sustained oscillations in \textbf{i}, decay to an arrested state in \textbf{f}, suppression of global CM transport despite persistent internal circulation in the vortex-like state \textbf{g}, and sustained orbital motion in the stable-loop state \textbf{h}.
	}	
		\label{figure02}
	\end{figure*}
	
	\subsection*{Self-alignment under anisotropic confinement}
	
	Velocity--polarity feedback provides a generic route by which local mechanical response can be converted into autonomous active motion. To isolate this mechanism and its coupling to geometry at the many-body level, we consider a minimal two-dimensional fluid of $N$ interacting self-aligning active Brownian particles, illustrated schematically in Fig.~\ref{figure01}. Particle $i$ is described by its position $\mathbf r_i=(x_i,y_i)$ and propulsion direction $\hat{\mathbf n}_i=(\cos\theta_i,\sin\theta_i)$, and self-propels with a typical speed $v_0$. Translational and orientational fluctuations are characterised by the diffusivities $D_t$ and $D_r$, respectively, and $\mu$ denotes the translational mobility.
	Conservative forces, $\mathbf F_i=-\nabla_i U$, derive from the total potential $U(\mathbf r^N)=\sum_i V_{\rm ext}(\mathbf r_i)+\sum_{i<j}u(r_{ij})$, which is the sum of an external potential $V_{\rm ext}$ and a pairwise repulsive contribution $u(r_{ij})$ between particles, with $r_{ij}=|\mathbf r_i-\mathbf r_j|$. We choose $V_{\rm ext}$ as a smooth anisotropic harmonic confinement, $V_{\rm ext}(x,y)=k_x(x^2+\alpha y^2)/2$, where $k_x$ sets the overall strength of the trap and $\alpha=k_y/k_x$ controls its degree of anisotropy. We choose $u(r)$ to be a bounded Gaussian repulsion, $u(r)=\epsilon\exp(-r^2/\sigma^2)$, where $\epsilon$ and $\sigma$ set the interaction strength and range, respectively.
Thus $\alpha=1$ corresponds to isotropic confinement, whereas increasing $\alpha$ selectively stiffens the $y$ direction without introducing hard boundaries or geometric singularities.
	

	
	Several active systems, such as cells and active granular particles, exhibit a local feedback between translational motion and propulsion orientation. By introducing the particle velocity $\mathbf v_i =\dot{\mathbf{r}}_i$, this mechanism can be modeled through an effective torque, termed self-aligning torque, $\mathcal T_i^{\rm sa}$. This torque aligns the propulsion direction with the instantaneous direction of motion, producing an effective angular drift $\omega_i^{\rm sa} = \mathcal T_i^{\rm sa}/\gamma_r$, with $\gamma_r$ the rotational friction coefficient, given by
	\begin{equation}
		\omega_i^{\rm sa}
		=
		\lambda
		\left(
		\hat{\mathbf n}_i\times\mathbf v_i
		\right)\cdot\hat{\mathbf e}_z \,.
		\label{eq:self_alignment_model}
	\end{equation}
Here, $\lambda$ is the self-alignment coefficient, $\hat{\mathbf e}_z$ is the unit vector normal to the plane, and $\lambda=\beta/\gamma_r$, where $\beta>0$ denotes the strength of the self-alignment torque. The coefficient $\lambda$ corresponds to the inverse of the self-alignment length, i.e., $\lambda^{-1}$ is the typical distance travelled by a particle while its orientation aligns with its velocity. Since $v_0\hat{\mathbf n}_i$ is parallel to $\hat{\mathbf n}_i$, self-propulsion by itself produces no aligning torque; reorientation is generated by the transverse motion induced by confinement and interparticle forces. Self-alignment therefore converts mechanical deflection into orientational response. Its dimensionless strength is $B=\lambda\sigma=\beta\sigma/\gamma_r$, which compares the interaction range $\sigma$ with the self-alignment length $\lambda^{-1}$. Together, $B$ and $\alpha$ form the two principal control parameters of the system.

	We investigate this microscopic dynamics from both particle and field perspectives. Brownian-dynamics (BD) simulations directly resolve the stochastic $N$-particle equations and therefore provide a particle-resolved realization of the model. We also develop a continuum theory derived from the same microscopic dynamics, allowing the particle-level motion to be recast in terms of collective density, orientational and current fields. Details of the BD implementation are given in the Methods.
	
To construct the continuum description, we formulate, to our knowledge, a dynamical density-functional theory (DDFT) for self-aligning active particles. DDFT provides a microscopic route from interacting Brownian dynamics to time-dependent one-body fields and has been developed for a broad range of equilibrium and nonequilibrium settings, including driven colloidal systems, active microswimmers and strongly inhomogeneous transport processes \cite{ArcherEvans2004,KrinningerBrader2016,Menzel2016,he2021dynamical, Wassermair2026PhaseDiagrams}. More recent developments have further extended density-based descriptions to unconventional transport with antisymmetric and chiral response \cite{kalz2022collisions, Abdoli_2026,Abdoli2026OddDDFT,avni2026odd}  without including self-aligning mechanisms. This additional torque introduces a qualitatively different ingredient into this framework: translational transport itself feeds back on the propulsion orientation, so that density transport and orientational dynamics become dynamically interlocked. 

We describe the system through the one-body position--orientation distribution function $\psi(\mathbf r,\theta,t)$, which gives the density of particles at time $t$ with position $\mathbf r$ and orientation $\theta$. 
By integrating the many-body Smoluchowski equation over the degrees of freedom of the remaining $N-1$ particles, we obtain the exact one-body conservation law, which governs the time evolution of $\psi(\mathbf r,\theta,t)$ (see the Methods):
	\begin{equation}
		\partial_t\psi
		=
		-\nabla\cdot\mathbf J
		-\partial_\theta J_\theta,
		\label{eq:main_ddft}
	\end{equation}
where $\mathbf J(\mathbf r,\theta,t)$ and $J_\theta(\mathbf r,\theta,t)$ are the translational and orientational currents in position--orientation space, respectively. 
The former accounts for the contributions of external forces, interactions, thermal fluctuations, and self-propulsion, while the latter accounts for rotational diffusion and the self-aligning torque.
Equation~\eqref{eq:main_ddft} is exact; the theoretical approximation enters when these currents are expressed solely in terms of one-body quantities.
	
	To close the one-body dynamics we express the currents in terms of average one-body fields, such as the number density $\rho(\mathbf r,t)=\int_0^{2\pi}d\theta\,\psi$ and the translational density current $\mathbf J_\rho(\mathbf r,t)=\int_0^{2\pi}d\theta\,\mathbf J$. This closure is obtained in three physically distinct steps.
	First, the instantaneous nonequilibrium interaction force is treated within the standard adiabatic approximation of DDFT: at each time, the interaction correlations are represented by those of an equilibrium reference system constrained to reproduce the instantaneous one-body density. Second, for the bounded Gaussian repulsion, the excess free energy of this reference system is evaluated at mean-field level. These two density-functional approximations determine the conservative interaction contribution to the translational current. Third, self-alignment introduces an additional active closure because the microscopic torque depends explicitly on the instantaneous particle velocity{\color{blue}.} At the one-body level, we represent this velocity by the local transport velocity $\mathbf v_{\rm loc}(\mathbf r,t)=\mathbf J_\rho/\rho$. This last closure is the key ingredient of our theory, which allows us to reproduce the simulation results theoretically.
	Within this closed description, the translational and orientational currents are
	\begin{subequations}
	\begin{align}
		\mathbf J
		&=
		v_0\hat{\mathbf n}\psi
		-
		\mu\psi\nabla
		\frac{\delta\mathcal F[\psi]}{\delta\psi},
		\label{eq:ddft_trans_current}
		\\
		J_\theta
		&=
		\lambda
		\left[
		\hat{\mathbf n}(\theta)\times\mathbf v_{\rm loc}(\mathbf r,t)
		\right]\cdot\hat{\mathbf e}_z\,\psi
		-
		D_r\partial_\theta\psi.
		\label{eq:ddft_rot_current}
	\end{align}
	\end{subequations}
	Here, $\mu$ is the particle mobility, with $\gamma=\mu^{-1}$ the corresponding translational friction coefficient, and it satisfies the Einstein relation $D_t=\mu k_{\rm B}T$, with $k_{\rm B}$ the Boltzmann constant and $T$ the temperature. 
	The structure of these currents makes the central feedback explicit. Propulsion, confinement and interactions generate translational transport; the resulting local velocity rotates the propulsion field through self-alignment; and the modified orientational distribution in turn reconstructs the active translational current. The theory therefore closes a dynamical current--orientation feedback loop rather than prescribing an effective aligning field externally.  Rotational diffusion enters through the angular gradient term of $J_\theta$, while the self-aligning contribution cannot be derived from an equilibrium free-energy functional, and is thus intrinsically a non-equilibrium term. Details of the DDFT derivation are given in Methods.
	
	The conservative part of the dynamics is encoded in the Helmholtz free-energy functional $\mathcal F=\mathcal F_{\rm id}+\mathcal F_{\rm ext}+\mathcal F_{\rm exc}$. The three contributions have transparent physical roles: the ideal term $\mathcal F_{\rm id}$ describes entropic spreading due to thermal fluctuations, the external term $\mathcal F_{\rm ext}$ encodes the anisotropic confinement, and the excess term $\mathcal F_{\rm exc}$ accounts for interparticle repulsion. For the present ultrasoft fluid, the excess term is evaluated at mean-field level, $\mathcal F_{\rm exc}\simeq\mathcal F_{\rm exc}^{\rm MF}$, and the three terms read
	\begin{subequations}
	\begin{align}
				\label{eq:free_energy_main_a}
		\mathcal F_{\rm id}[\psi]
		&=
		k_{\rm B}T
		\int d\mathbf r\,d\theta\,
		\psi
		\left[
		\ln\!\left(\Lambda^2\psi\right)-1
		\right],
		\\
		\mathcal F_{\rm ext}[\rho]
		&=
		\int d\mathbf r\,
		\rho(\mathbf r)V_{\rm ext}(\mathbf r),
		\\
		\mathcal F_{\rm exc}^{\rm MF}[\rho]
		&=
		\frac{1}{2}
		\int d\mathbf r\,d\mathbf r'\,
		\rho(\mathbf r)
		u(|\mathbf r-\mathbf r'|)
		\rho(\mathbf r').
		\label{eq:free_energy_main}
	\end{align}
	\end{subequations}
	Here $\Lambda$ is an arbitrary microscopic reference length whose value drops out of the gradients governing the dynamics. Note that the adiabatic and mean-field approximations enter at different levels: the former maps the instantaneous nonequilibrium interaction force onto an equilibrium reference problem, whereas the latter specifies the excess functional used to represent that reference state. In this way, conservative forces are treated through the density-functional free energy, while self-alignment remains an explicitly active contribution that cannot be derived from $\mathcal F$. The microscopic derivation and the corresponding force- and torque-balance relations are given in Methods.
	
	In both the particle-resolved simulations and the theory, the self-aligning active fluid is initialized as a localized density cloud displaced from the trap centre [Fig.~\ref{figure01}(c)] with an isotropic orientational distribution. This preparation imposes neither an initial polarization nor a prescribed orbital pathway, so that any collective motion emerges spontaneously. 

\begin{figure*}[t!]
\centering
\includegraphics[width=0.9\textwidth]{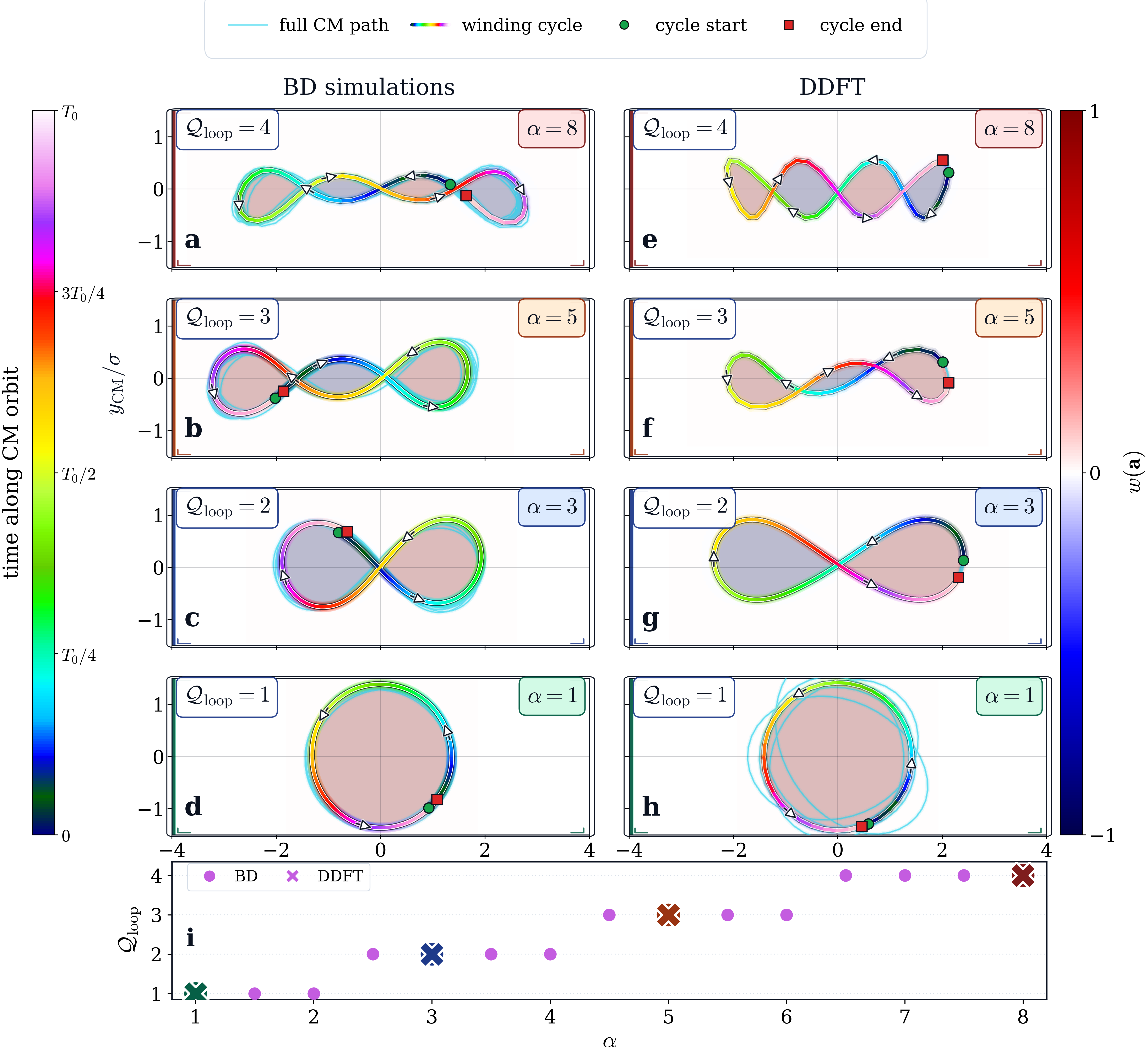}
\caption{ \textbf{Orbit topology of the collective centre-of-mass motion.} Representative late-time centre-of-mass (CM) trajectories obtained from \textbf{(a--d)} Brownian-dynamics (BD) simulations and \textbf{(e--h)} dynamical density functional theory (DDFT), shown for decreasing trap anisotropy $\alpha=8,5,3,$ and $1$ from top to bottom, where $\mathrm{Pe}=v_0/(D_r\sigma)=50$, $\widetilde{D}_t=D_t/(D_r\sigma^2)=0.1$, $\widetilde{\epsilon}=\epsilon/(\gamma D_r\sigma^2)=500$, and $B=10$. The coloured segment denotes one fundamental orbital period $T_0$; colour marks the progression of time and arrows indicate the direction of traversal. The faint cyan curve shows the retained late-time trajectory. The background is the winding field $w(\mathbf a)$ of the selected orbit, with positive (red) and negative (blue) domains denoting counter-clockwise and clockwise encirclement, respectively. The connected non-zero winding domains define the orbital loop number $\mathcal Q_{\rm loop}$. Our results indicate a discrete hierarchy $\mathcal Q_{\rm loop}=1,2,3,$ and $4$ as the anisotropy of the confinement is continuously increased, as shown in (\textbf{i}).  
}
\label{fig:orbital_topology}
\end{figure*}

\subsection*{Collective state selection by self-alignment and confinement}

A self-aligning confined active fluid organises into four robust steady dynamical states, depending on the reduced strength of self-alignment $B$ and the potential asymmetry $\alpha$ [Fig.~\ref{figure02}(a)]. 
These states fall into two broad classes depending on whether the cluster as a whole is transported. In the first class the cluster stays nearly fixed: it either comes to rest (\emph{arrested} state; see Supplementary Video~1) or rotates on itself, with its constituents circulating around an almost stationary centre of mass (\emph{vortex-like} state; see Supplementary Video~2). In the second class the cluster moves persistently: it either shuttles back and forth without completing a closed orbit (\emph{oscillatory} state; see Supplementary Video~4) or travels along a closed orbit (\emph{stable-loop} state; see Supplementary Video~3). Closed orbits are not all equivalent: they come in distinct classes (a simple loop, a figure-eight, multi-lobed paths) labelled by an integer winding-based loop number, which is the central object of this work and is introduced below.  The diagram should be interpreted as a nonequilibrium dynamical-state map rather than an equilibrium phase diagram: the shaded boundaries mark crossovers between the responses resolved on the sampled $(B,\alpha)$ grid.

The identification of these states arises from monitoring the steady-state centre of mass dynamics, looking at its position $\mathbf R_{\rm CM}(t)$ and velocity $\mathbf V_{\rm CM}(t)=\dot{\mathbf R}_{\rm CM}(t)$ (see Methods for the explicit definitions).
A direct measure of the collective response (dynamical state) is the centre of mass velocity normalised by the propulsion speed,
\begin{equation}
	\mathbf v_{\rm CM}(t)
	=
	\frac{\mathbf V_{\rm CM}(t)}{v_0}.
	\label{eq:cm_velocity_dimensionless}
\end{equation}
This observable isolates transport of the collective coordinate from internal rearrangements of the active cloud. Its magnitude $v_{\rm CM}(t)$ and Cartesian components first separate the four states into two pairs, according to whether the centre of mass is transported in the steady state.

Two states have suppressed centre of mass transport. In the \emph{arrested} state, both velocity components decay and $v_{\rm CM}\to0$: the collective coordinate comes to rest [Fig.~\ref{figure02}(f)]. A nearly stationary centre of mass, however, does not imply that the active material is at rest. In the \emph{vortex-like} state, global centre of mass transport is strongly suppressed while the particle cloud sustains pronounced internal circulation around an almost fixed centre of mass [Fig.~\ref{figure02}(g)]. The two states are thus distinguished not by $v_{\rm CM}$, which is small in both, but by the internal flow, quantified below through vorticity and density-weighted enstrophy.

The other two states have sustained centre of mass transport, so that their centre of mass speed remains finite and its velocity exhibits sustained, periodic modulations. However, in the \emph{oscillatory} state, the late-time trajectory $\mathbf R_{\rm CM}(t)$ does not converge to a robust closed orbit [Fig.~\ref{figure02}(i)], at variance with the \emph{stable-loop} state, where the periodic response of the centre of mass velocity generate a closed trajectory [Fig.~\ref{figure02}(h)]. 
In other words, only the stable-loop state has a well-defined closed orbit, and hence a well-defined orbital topology, which is analyzed in the next section.

The dynamical-state map reveals a non-monotonic, re-entrant dependence of the long-time dynamics on the control parameters. Changing self-alignment $B$ at fixed confinement $\alpha$ can reorganise the system between arrested, vortex-like, stable orbiting, and oscillatory dynamics, and back to an arrested state, as for $\alpha=5$, while changing $\alpha$ at fixed $B$ can select a different collective state. This sequence can be understood as a competition between the alignment torque, which turns each polarity toward the force acting on the particle, and rotational diffusion, which randomizes it. The force has two parts: the confining force, which is nearly the same for all particles, and the repulsion from neighbours, which differs from particle to particle and fluctuates rapidly. 

At small $B$, alignment cannot overcome rotational diffusion, polarities stay disordered and the cluster is arrested. At intermediate $B$, particles align mainly with their own interaction force and circulate around the trap, with the interactions locking their senses of rotation; this gives vortex-like states with large internal angular momentum but little motion of the centre of mass. At larger $B$, alignment is fast enough for the polarities to lock onto the common confining force, so that the cluster moves coherently and its centre of mass follows a closed loop. At even larger $B$, we attribute the return to an arrested state to polarities following the local, particle-specific force too quickly to remain coherent with one another, which suppresses the net polarization. Anisotropic confinement therefore plays a dynamical role beyond setting the accessible spatial extent: together with local self-alignment, it selects how microscopic polarity--motion feedback is converted into global collective motion. Within the stable-loop region, orbits with comparable speeds and characteristic frequencies can differ in the number and orientation of their lobes, and this difference is captured by an integer loop number, built from the winding of the centre-of-mass trajectory. This suggests a topological classification of the orbits, which we develop in the next section.

\begin{figure*}[t] 
	\centering \includegraphics[width=\textwidth]{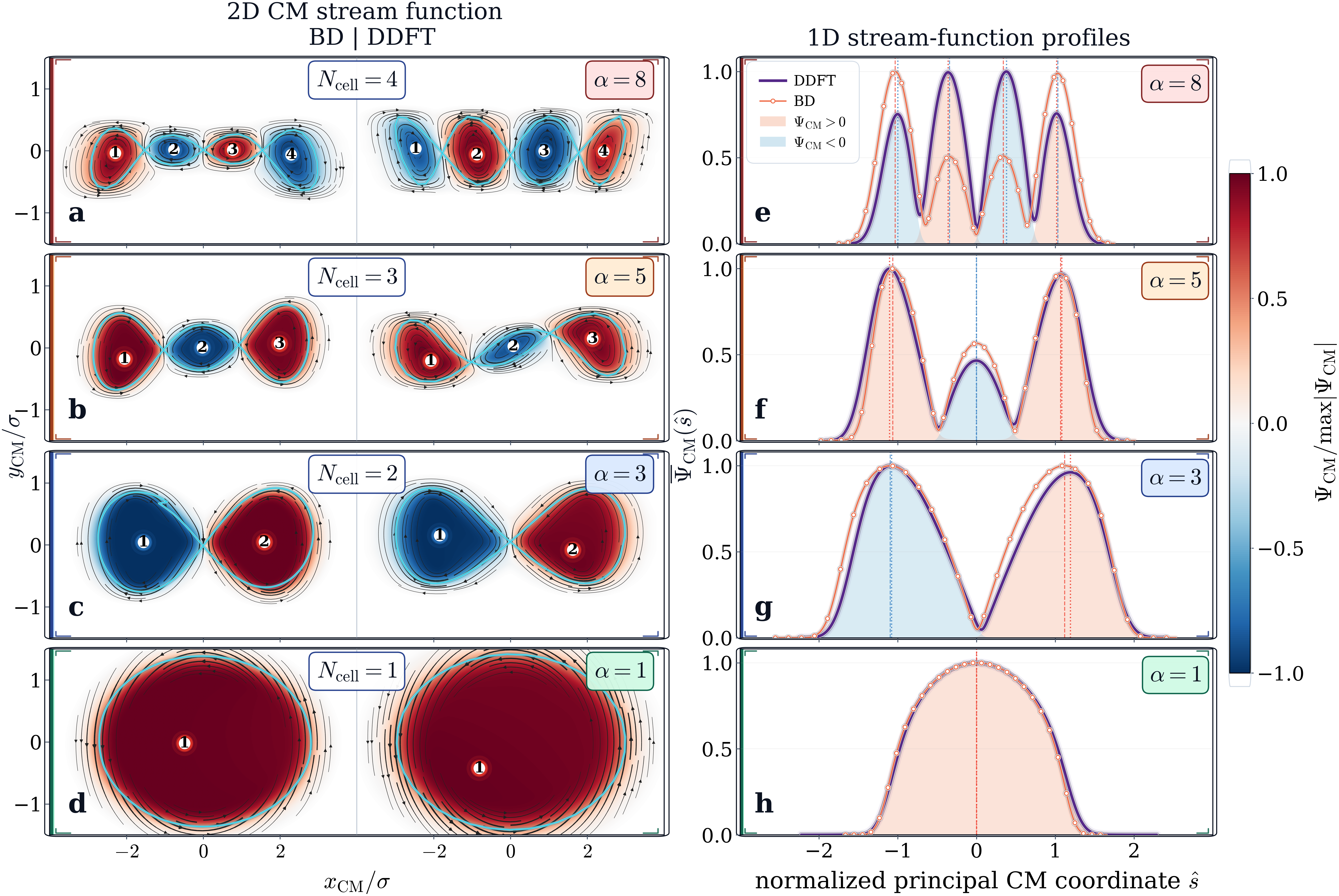}
	\caption{ \textbf{Recirculating probability-current structure associated with the selected orbital topologies.} The same sequence $\alpha=8,5,3,$ and $1$ is shown from top to bottom. \textbf{(a--d)} Normalised CM stream function $\Psi_{\rm CM}(x_{\rm CM},x_{\rm CM})$ reconstructed from the late-time CM probability current, with BD on the left and DDFT on the right of each panel. Black streamlines show the solenoidal probability flow; red and blue regions denote opposite circulation senses. Numbered markers indicate the centres of the recirculating cells. \textbf{(e--h)} Normalised longitudinal circulation profiles $\overline{\Psi}_{\rm CM}(\hat s)$ obtained by integrating the magnitude of the stream function across the transverse principal direction. Purple curves show DDFT and orange symbol--line curves show BD; red and blue shaded contributions indicate the two circulation senses in the DDFT field. The probability-current reconstruction yields $N_{\rm cell}=4,3,2,$ and $1$, matching the loop numbers in Fig.~\ref{fig:orbital_topology} obtained from the winding analysis of the same data.
	} \label{fig:cm_streamfunction}
\end{figure*}


\subsection*{Orbit topology and probability-current circulation}

Self-alignment and confinement do more than determine whether the active cloud moves coherently: within the stable-loop regime they select the winding structure and thus the \emph{topology} of the collective trajectory. By changing the potential asymmetry $\alpha$ and self-alignment $B$, we can obtain a simple orbit [Fig.~\ref{fig:orbital_topology}(d)], a figure-eight [Fig.~\ref{fig:orbital_topology}(c)], and even higher-order multi-lobed paths [Fig.~\ref{fig:orbital_topology}(a--b)], with comparable spatial and temporal scales while differing in how they wind around the plane. Specifically, $B$ controls the stability of the orbiting regime, while geometry primarily selects the orbital topology.

We characterise the closed late-time centre-of-mass trajectory by its winding about points in the plane. For one fundamental cycle $\Gamma$ of $\mathbf R_{\rm CM}(t)=(x_{\rm CM},y_{\rm CM})$, of period $T_0$, and for a generic point $\mathbf a=(a_x,a_y)$ of the centre-of-mass plane not lying on $\Gamma$, the winding number of $\Gamma$ around $\mathbf a$ is
	\begin{equation}
		w(\mathbf a)
		\!=\!
		\frac{1}{2\pi}
		\oint_{\Gamma}
		\frac{
			(x_{\rm CM}-a_x)\dot y_{\rm CM}
			-(y_{\rm CM}-a_y)\dot x_{\rm CM}
		}{
			(x_{\rm CM}-a_x)^2+(y_{\rm CM}-a_y)^2
		}dt ,
		\label{eq:winding_field}
	\end{equation}
	 i.e. the net angle swept by the vector $\mathbf R_{\rm CM}(t)-\mathbf a$ in one period, in units of $2\pi$. Here $w=+1$ and $w=-1$ denote one counter-clockwise and clockwise encirclement, respectively. Self-intersections divide the centre-of-mass plane into connected domains $\mathcal D_\ell$ of constant integer winding $w_\ell$ (the unbounded outer domain has $w=0$). We define
	\begin{equation}
		\mathcal Q_{\rm loop}
		=
		\sum_{\mathcal D_\ell}|w_\ell| ,
		\label{eq:orbital_loop_number}
	\end{equation}
so that lobes traversed in opposite directions contribute independently rather than cancelling. Thus $\mathcal Q_{\rm loop}=1$ for a simple closed orbit and $\mathcal Q_{\rm loop}=2$ for a figure-eight, while larger values resolve progressively more complex multi-lobed trajectories. For all orbits considered here every non-zero domain has $|w_\ell|=1$, so that $\mathcal Q_{\rm loop}$ coincides with the number of lobes.

Figure~\ref{fig:orbital_topology} reveals a discrete hierarchy $\mathcal Q_{\rm loop}=1,2,3,$ and $4$ as the confinement becomes more anisotropic. The transition between these states is not a gradual change of orbit size or frequency, but a reorganisation of the winding-domain structure of the collective trajectory itself. Increasing continuously the potential anisotropy $\alpha$ therefore selects distinct orbital sectors with different numbers and orientations of connected winding regions, establishing a genuinely discrete topological organisation of the cluster motion (see Fig.~\ref{fig:orbital_topology}(i)).

The winding field is built from a single coherent closed orbit, and is therefore best suited to clean, nearly periodic trajectories. We thus ask whether the same structure is also encoded statistically in the late-time probability flow of the centre of mass, which is accessible also when individual orbits are noisy. To this aim, we introduce the time-averaged probability density $P_{\rm CM}(\mathbf R)$ of finding the centre of mass at $\mathbf R$ and the corresponding probability current $\mathbf J_{\rm CM}(\mathbf R)$, which also records the direction of motion at each position:
\begin{subequations}
	\begin{align}
		P_{\rm CM}(\mathbf R)
		&=
		\Big\langle
		\delta\!\left[\mathbf R-\mathbf R_{\rm CM}(t)\right]
		\Big\rangle_t, \label{eq:cm_probability}
		\\
		\mathbf J_{\rm CM}(\mathbf R)
		&=
		\Big\langle
		\mathbf V_{\rm CM}(t)\,
		\delta\!\left[\mathbf R-\mathbf R_{\rm CM}(t)\right]
		\Big\rangle_t .
		\label{eq:cm_probability_current}
	\end{align}
\end{subequations}
Here $\langle\cdots\rangle_t$ is the average over the late-time trajectory. A non-zero $\mathbf J_{\rm CM}$ in the stationary state signals sustained circulation of the CM, which is forbidden at equilibrium by detailed balance.

The circulating part of $\mathbf J_{\rm CM}$ is represented by a scalar stream function $\Psi_{\rm CM}$ through $\mathbf{J}^{\rm sol}_{\rm CM}
=
(\partial_{y_{\rm CM}}\Psi_{\rm CM},
-\partial_{x_{\rm CM}}\Psi_{\rm CM})$. Positive and negative values of $\Psi_{\rm CM}$ correspond to counter-clockwise and clockwise circulation, respectively. Each coherent recirculating region of the centre-of-mass current, which we call a \emph{cell}, is a connected region of the centre-of-mass plane containing a local extremum of $\Psi_{\rm CM}$. The number of cells is
	\begin{equation}
		N_{\rm cell}
		=
		\#\big\{\text{connected extrema of }\Psi_{\rm CM}\big\},
		\label{eq:ncell_definition}
	\end{equation}
where maxima and minima are counted together, as they correspond to opposite circulation senses. The reconstruction procedure and the normalised profile of $\Psi_{\rm CM}$ along the principal centre-of-mass direction, used to compare BD and DDFT in Fig.~\ref{fig:cm_streamfunction}(e--h), are given in Methods. This projection makes the number and relative placement of the cells directly comparable even when BD and DDFT differ slightly in the spatial extent of their centre-of-mass distributions.

Across the coherent loop states shown in Figs.~\ref{fig:orbital_topology},  the independently reconstructed probability flow reproduces the same hierarchy as the trajectory winding,
	\begin{equation}
		N_{\rm cell}=\mathcal Q_{\rm loop}.
		\label{eq:cell_loop_correspondence}
	\end{equation}
For a noiseless periodic orbit this correspondence is expected, since $\mathbf J_{\rm CM}$ reduces to a line current along $\Gamma$ and $\Psi_{\rm CM}=w/T_0$. The nontrivial finding is that it survives thermal noise and coarse-graining in both BD and DDFT: $\mathcal Q_{\rm loop}$ is computed from a single closed trajectory, whereas $N_{\rm cell}$ is extracted
from a coarse-grained probability-flow field. Their equality shows that the
selected orbital structure leaves a directly resolvable cellular signature in
the nonequilibrium circulation of the collective coordinate.

\subsection*{Internal vortex architecture and rotational activity}

\begin{figure*}[t]
	\centering
	\includegraphics[width=\textwidth]{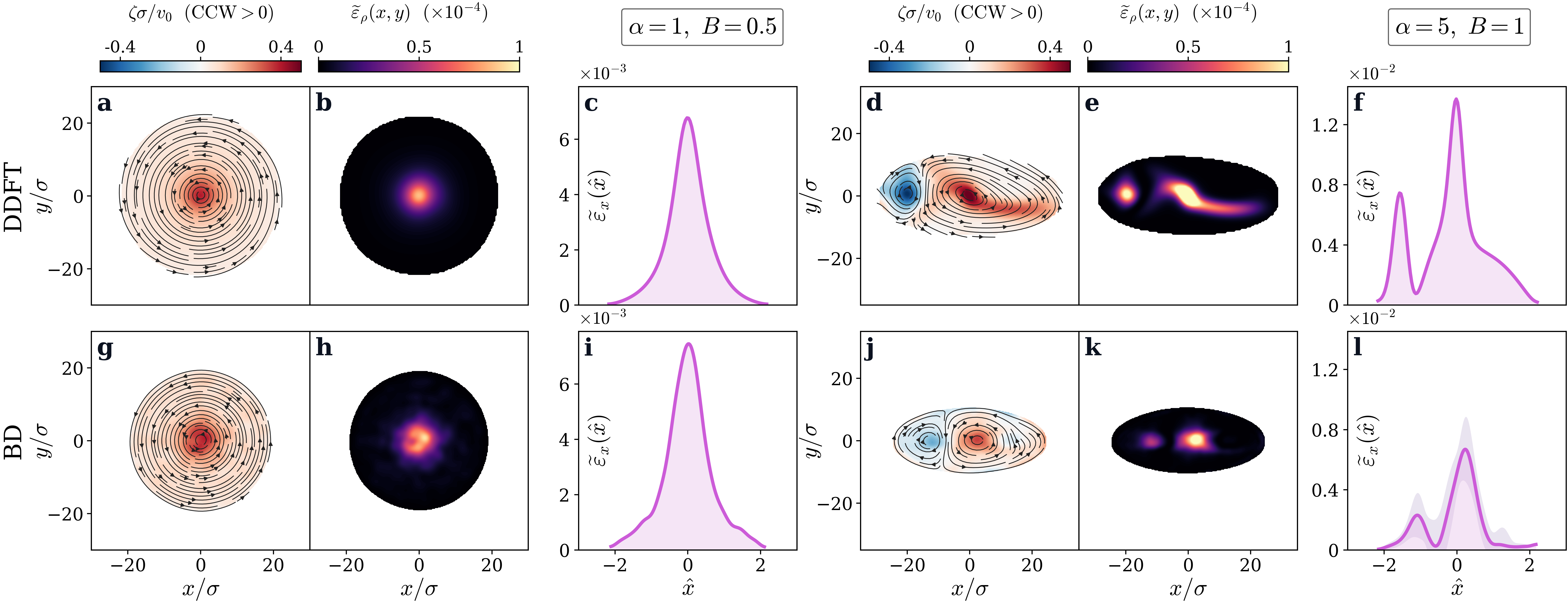}
	\caption{\textbf{Internal vortex architecture and localization of rotational activity in DDFT and BD.}
	The upper row (\textbf{a--f}) shows DDFT results and the lower row (\textbf{g--l}) the corresponding Brownian-dynamics (BD) results. The left group, \textbf{a--c,g--i}, corresponds to $\alpha=1$ and $B=0.5$, whereas the right group, \textbf{d--f,j--l}, corresponds to $\alpha=5$ and $B=1$. \textbf{a,d,g,j}, Vorticity $\zeta\sigma/v_0$ of the internal velocity field in the centre-of-mass frame, with positive (red) and negative (blue) values denoting counter-clockwise and clockwise local rotation, respectively. Black streamlines show the corresponding internal flow $\mathbf v_{\rm int}$. \textbf{b,e,h,k}, Density-weighted enstrophy $\widetilde{\varepsilon}_{\rho}$, which identifies where rotational activity is concentrated irrespective of its circulation sense. \textbf{c,f,i,l}, Corresponding longitudinal enstrophy profile $\widetilde{\varepsilon}_{x}(\hat{x})$, obtained by integrating the local enstrophy across the transverse direction and expressing position
	in units of the longitudinal cloud width; the normalisation is defined
	in Methods. The isotropic state ($\alpha=1$) supports a single coherent counter-clockwise vortex with centrally localized rotational activity, whereas anisotropic confinement reorganises the flow into neighbouring counter-rotating cells with spatially separated enstrophy maxima. For
	the $\alpha=5$, $B=1$ BD state, the solid curve is the average over
	realizations that settle into the two-cell vortex state and the shaded
	region in \textbf{l} denotes the corresponding standard error of the
	mean; the $\alpha=1$, $B=0.5$ BD panels are obtained from a representative single realization.}
		\label{fig:vortex_enstrophy}
	\end{figure*}
	
	
	The centre-of-mass orbit describes how the collective coordinate moves through space, but it does not specify how the active material moves relative to that
coordinate. This distinction is essential for the vortex-like states identified above: weak global transport can coexist with sustained internal circulation. We therefore resolve the real-space flow in the instantaneous centre-of-mass frame and use it to characterise a second, complementary level of organisation: the vortex architecture within the active cloud.
	
With $\mathbf r_{\rm c}=\mathbf r-\mathbf R_{\rm CM}(t)$, the internal velocity field is
	\begin{equation}
		\mathbf v_{\rm int}(\mathbf r_{\rm c},t)
		=
		\frac{\mathbf J_{\rho}(\mathbf r_{\rm c},t)}
		{\rho(\mathbf r_{\rm c},t)}
		-
		\mathbf V_{\rm CM}(t),
		\label{eq:internal_velocity}
	\end{equation}
where $\rho$ is the number density and $\mathbf J_{\rho}$ is the orientationally integrated translational current, both evaluated in the co-moving frame. For BD, the corresponding density and current fields are obtained by
coarse-graining particle positions and velocities; for DDFT they follow
directly from the one-body density and current. In the following, $x$ and $y$
denote coordinates in this co-moving frame.  Since $\mathbf V_{\rm CM}$ is spatially uniform, its subtraction does not affect the vorticity defined below: it only removes the rigid translation of the cloud from the streamlines.
	
The local sense and strength of rotation are encoded by the out-of-plane
vorticity
	\begin{equation}
		\zeta(\mathbf r)
		=
		\bigl[\boldsymbol{\nabla}\times\mathbf v_{\rm int}(\mathbf r)\bigr]_z
		=
		\partial_x v_{{\rm int},y}
		-
		\partial_y v_{{\rm int},x}.
		\label{eq:internal_vorticity}
	\end{equation}
Thus $\zeta>0$ and $\zeta<0$ identify counter-clockwise and clockwise local
circulation, respectively. The vorticity maps in  Fig.~\ref{fig:vortex_enstrophy}(a,g,d,j) therefore reveal the signed vortical structure
of the internal flow rather than the winding of the centre-of-mass trajectory.
	
To quantify where that rotational motion is dynamically concentrated, we use
the density-weighted enstrophy density [colour gradient in Fig.~\ref{fig:vortex_enstrophy}(b,h,e,k)]
	\begin{equation}
		\widetilde{\varepsilon}_{\rho}(\mathbf r)
		=
		\frac{\sigma^2\rho(\mathbf r)}{2N}
		\left[
		\frac{\sigma\,\zeta(\mathbf r)}{v_0}
		\right]^2 .
		\label{eq:local_enstrophy_density}
	\end{equation}
Unlike vorticity, this field is positive definite and therefore measures the
local intensity of rotational activity without retaining its chirality. The
density weighting emphasizes circulation carried by the particle-rich part of
the cloud and suppresses spurious contributions from dilute tails. 

In Fig.~\ref{fig:vortex_enstrophy}(c,i,f,l), we integrate this
quantity across the transverse coordinate $y$, which is aligned with a principal axis of the potential, obtaining a one-dimensional profile  along the longitudinal coordinate $x$
	\begin{equation}
		\widetilde{\varepsilon}_{x}(\hat{x})
		=
		\frac{R_x}{\sigma}
		\int d\!\left(\frac{y}{\sigma}\right)
		\widetilde{\varepsilon}_{\rho}(x,y),
		\qquad
		\hat{x}=\frac{x-x_{\rm c}}{R_x},
		\label{eq:longitudinal_enstrophy}
	\end{equation}
where $x_{\rm c}$ and $R_x$ denote the longitudinal density centre and
root-mean-square cloud width, respectively; their numerical construction is
given in Methods. This representation separates changes in the spatial
organisation of rotational activity from small differences in the overall
cloud size between BD and DDFT.
	
Figure~\ref{fig:vortex_enstrophy} exposes a confinement-driven reorganisation of the internal flow.  For isotropic confinement in the vortex phase (for instance, $\alpha=1$, $B=0.5$) both DDFT and BD show a single coherent counter-clockwise vortex centred on the cloud [Figs.~\ref{fig:vortex_enstrophy}(a,g)]. Its rotational activity is localized near the centre, producing a single dominant enstrophy maximum [Figs.~\ref{fig:vortex_enstrophy}(b,h) and Figs.~\ref{fig:vortex_enstrophy}(c,i)]. 
 The agreement between BD and DDFT shows that the continuum theory captures not only the direction of internal circulation but also where within the active fluid that circulation is concentrated.
	
Increasing the anisotropy to $\alpha=5$ at $B=1$ does more than elongate this flow pattern. The internal current reorganises into neighbouring  vortices of opposite vorticity [Figs.~\ref{fig:vortex_enstrophy}(d,j)],  showing that confinement can split a single coherent circulation into a counter-rotating vortex architecture. Correspondingly, the density-weighted enstrophy develops spatially separated regions of strong rotational activity [Figs.~\ref{fig:vortex_enstrophy}(e,k)], and its longitudinal projection resolves these regions as distinct maxima [Figs.~\ref{fig:vortex_enstrophy}(f,l)].  For BD, the profile in Fig.~\ref{fig:vortex_enstrophy}(l) is averaged over the realizations that reach this two-vortex state. Anisotropic confinement therefore controls not only the global collective trajectory but also the internal partitioning of rotational motion within the active fluid.
	
	
	Together with the orbital and  centre-of mass current analyses, these results establish a hierarchy of nonequilibrium organisation. The loop number $\mathcal Q_{\rm loop}$ characterises the topology of the collective coordinate, $N_{\rm cell}$ resolves circulation in the centre-of-mass state space, and $\zeta(\mathbf r)$ together with $\widetilde{\varepsilon}_{\rho}(\mathbf r)$ resolves circulation within the material itself.

	\section*{Discussion}
	
	Our results establish the self-sustained trajectory of a finite active assembly as a topological object. In the minimal realization considered here, local self-alignment converts mechanically induced deflections into changes of propulsion, while anisotropic confinement controls how this feedback is organised in space. The same interacting cluster can thereby access qualitatively different long-time dynamics and, within the orbiting regime, spontaneously select a discrete hierarchy of topologically distinct closed trajectories. The associated orbital loop number $\mathcal Q_{\rm loop}$ belongs to the collective centre-of-mass path itself rather than to an imposed geometric structure or an internal orientational field. Earlier studies established orbital motion and multiple confined states at the single-particle level \cite{Dauchot2019,Damascena2022,Damascena2023}, as well as collective orbiting and circling states in interacting self-aligning systems \cite{Canavello2024,Musacchio2026Circling} or   non-aligning chiral particles \cite{caprini2024self}. Here the many-body collective coordinate itself carries the winding-domain structure, making orbit topology an emergent property of the active cluster. The recovery of the same hierarchy in particle-resolved dynamics further shows that this organisation is a property of the interacting active assembly rather than of the continuum representation used to describe it.
	
	
	A second central result is that this trajectory-level topology is reproduced in the nonequilibrium probability flow of the collective coordinate. The orbital loop number $\mathcal Q_{\rm loop}$ is obtained from the connected winding domains of a single coherent cycle, whereas $N_{\rm cell}$ is extracted from a coarse-grained centre-of-mass probability current. Their observed correspondence, $N_{\rm cell}=\mathcal Q_{\rm loop}$, therefore links two descriptions of the same collective dynamics. The equality reveals that the selected orbit topology leaves a directly resolvable signature in probability space. The current-based viewpoint also connects naturally to broader efforts to characterise driven soft matter through spatially resolved nonequilibrium transport and flow organisation \cite{SvetlizkyRoichman2021,avni2026hall}. That a density-based description reproduces both the orbital hierarchy and its probability-current structure also illustrates how nonequilibrium density-functional approaches can resolve dynamical state selection in interacting confined systems \cite{KrinningerBrader2016,Wassermair2026PhaseDiagrams}.

	
	The collective trajectory, however, does not uniquely determine how the active material moves internally. A cluster following a simple closed orbit may support a single coherent vortex, whereas stronger anisotropy can split the internal circulation into neighbouring counter-rotating regions; conversely, strong internal rotation can persist even when global centre-of-mass motion is weak. Orbit topology, probability-space circulation and real-space vortex organisation therefore constitute distinct levels of the same active-cluster dynamics. This separation is physically important: the global path describes how the assembly moves through space, whereas vorticity and density-weighted enstrophy resolve how activity is redistributed within the moving assembly. Structured environments are already known to reorganise collective loops and transport in driven particle systems \cite{Libal2020TimeCrystal,Volpe2016Disorder}; the present results show that smooth geometry alone, when coupled to local active feedback, can simultaneously organise the trajectory of the cluster and the circulation sustained inside it.
	
	The broader implication is that orbit topology need not be tied to the microscopic origin of the active motion. Self-alignment and anisotropic confinement provide a particularly transparent realization because mechanical forces are fed directly back into propulsion, but the classification itself is constructed solely from the collective trajectory. This distinction opens the possibility of identifying the same topological organisation in physically very different active assemblies. Mechanical polarity--motion feedback can selectively actuate active elastic networks \cite{Baconnier2022ActiveSolids} and generate autonomous cooperative transport in robotic swarms \cite{Arbel2025Transport,Fersula2026Phototaxis}; geometry and polarity likewise organise migration, coherent rotation and orbital motion in cellular assemblies and tissues \cite{Jain2020CellMigration,Brandstatter2023SphericalTissues,Kim2026Spheroids,alert2020physical}. These systems should not be regarded as direct realizations of the present microscopic model, but they illustrate the broader physical setting in which local active feedback and environmental geometry jointly determine collective paths. In this sense, orbit topology offers a common descriptor of autonomous cluster motion without requiring the trajectory itself to be prescribed.
	
	Several directions follow naturally from this picture. Competing self- and anti-self-alignment, known to reorganise clustering and collective order in dense active systems \cite{Musacchio2025MIPS}, may generate coexistence, frustration or switching between orbital sectors. Intrinsic chirality provides another route to recurrent collective motion \cite{Musacchio2026Circling}, raising the question of how chirality and geometric selection combine when both are present. Replacing the smooth harmonic confinement by constrictions, obstacle arrays or porous environments will test the robustness of the loop hierarchy in strongly structured landscapes, where confinement can profoundly reorganise both transport and many-body structure \cite{Sinaasappel2025Porous,Volpe2016Disorder,brader2026hard}. Noise may induce transitions between distinct $\mathcal Q_{\rm loop}$ attractors, in analogy with noise-driven switching between confined single-particle orbital states \cite{Damascena2023}, while hydrodynamic interactions and three-dimensional confinement may couple orbit topology to additional collective modes. More fundamentally, identifying the dynamical conditions under which orbital sectors are created, destroyed or exchanged should reveal whether changes of $\mathcal Q_{\rm loop}$ can be understood as bifurcations of the underlying current--orientation feedback. Defined solely by the collective trajectory, orbit topology transcends the microscopic details of the active mechanism and points towards a universal classification of recurrent motion in finite active assemblies.

	\section*{Methods}
	\label{Methods}

	\subsection*{Dynamical density-functional theory (DDFT)}
	\label{methods:ddft}
	Our continuum description is obtained by reducing the stochastic many-body
	dynamics of the self-aligning particles to a time-dependent one-body
	distribution in position--orientation space. The derivation follows the
	canonical DDFT strategy: we first formulate the exact many-body
	Smoluchowski dynamics, integrate out $N-1$ particles to obtain a one-body
	continuity equation, and then close the interaction force using the
	adiabatic approximation~\cite{rex2007dynamical}. For the ultrasoft Gaussian pair potential, the
	equilibrium excess free-energy functional entering this closure is evaluated
	at mean-field level. A second closure is required by self-alignment: the
	instantaneous particle velocity appearing in the microscopic aligning torque
	is represented at one-body level by the local transport velocity
	$\mathbf J_\rho/\rho$. This construction retains the feedback between
	particle currents and orientation that distinguishes self-aligning active
	matter from conventional active Brownian particles.
	
	\subsubsection*{One-body dynamics and density-functional closure}
	
	The DDFT is constructed from the same overdamped microscopic dynamics as
	the particle model. The position and orientation of particle $i$ evolve as
	\begin{align}
		\dot{\mathbf r}_i
		&=
		v_0\hat{\mathbf n}_i
		+
		\mu\mathbf F_i
		+
		\sqrt{2D_t}\,\boldsymbol\xi_i(t),
		\nonumber\\
		\dot{\theta}_i
		&=
		\lambda
		\left(
		\hat{\mathbf n}_i\times\mathbf v_i
		\right)\cdot\hat{\mathbf e}_z
		+
		\sqrt{2D_r}\,\eta_i(t),
		\label{eq:microscopic_dynamics}
	\end{align}
	where $\hat{\mathbf n}_i=(\cos\theta_i,\sin\theta_i)$ and
	$\mathbf r_i=v_0\hat{\mathbf n}_i+\mu\mathbf F_i$ denotes the
	deterministic translational drift. The conservative force is
	$\mathbf F_i=-\nabla_iV_{\rm ext}(\mathbf r_i)
	-\sum_{j\neq i}\nabla_i u(r_{ij})$.
	Here $\mu$ is the translational mobility, while $D_t$ and $D_r$ are the
	translational and rotational diffusion coefficients. The alignment
	coefficient is $\lambda=\beta/\gamma_r$, where $\beta$ is the microscopic
	self-alignment strength and $\gamma_r$ the rotational damping entering the
	alignment dynamics. The unit vector $\hat{\mathbf e}_z$ is normal to the
	two-dimensional plane. The noises $\boldsymbol\xi_i$ and $\eta_i$ are
	independent Gaussian white noises with zero mean and unit variance.
	
	Because the active contribution $v_0\hat{\mathbf n}_i$ is parallel to the
	orientation itself, it does not generate an aligning torque. The
	deterministic orientational drift can therefore equivalently be written as
	$\omega_i^{\rm sa}
	=\lambda\mu(\hat{\mathbf n}_i\times\mathbf F_i)\cdot\hat{\mathbf e}_z$.
	Self-alignment thus responds to the transverse component of the mechanical
	motion generated by confinement and interparticle forces.
	
	Let
	$P(\mathbf r^N,\theta^N,t)$ denote the $N$-body probability density, with
	$\mathbf r^N=(\mathbf r_1,\ldots,\mathbf r_N)$ and
	$\theta^N=(\theta_1,\ldots,\theta_N)$. The stochastic dynamics imply the
	many-body Smoluchowski equation
	\begin{align}
		\partial_tP
		={}&
		-\sum_{i=1}^{N}
		\nabla_i\cdot
		\left[
		\left(
		v_0\hat{\mathbf n}_i+\mu\mathbf F_i
		\right)P
		-
		D_t\nabla_iP
		\right]
		\nonumber\\
		&-
		\sum_{i=1}^{N}
		\partial_{\theta_i}
		\left[
		\omega_i^{\rm sa}P
		-
		D_r\partial_{\theta_i}P
		\right].
		\label{eq:nbody_smolu}
	\end{align}

	At the one-body level, the state of the system is described by the
	position--orientation density $\psi(\mathbf r,\theta,t)$, normalised such
	that $\int d\mathbf r\,d\theta\,\psi=N$. Its orientational integral gives
	the number density
	$\rho(\mathbf r,t)=\int_0^{2\pi}d\theta\,\psi(\mathbf r,\theta,t)$.
	Integrating the many-body Smoluchowski dynamics over all remaining degrees
	of freedom gives the exact conservation law
	\begin{equation}
		\partial_t\psi
		=
		-\nabla\cdot\mathbf J
		-
		\partial_\theta J_\theta .
		\label{eq:onebody_continuity_methods}
	\end{equation}
	Before closure, the translational current can be written as
	\begin{equation}
		\mathbf J
		=
		v_0\hat{\mathbf n}\psi
		-
		D_t\nabla\psi
		-
		\mu\psi\nabla V_{\rm ext}
		+
		\mu\mathbf F_{\rm int},
		\label{eq:exact_onebody_current}
	\end{equation}
	where
	$\mathbf F_{\rm int}
	=-\int d\mathbf r'\,d\theta'\,
	\psi^{(2)}(\mathbf r,\theta,\mathbf r',\theta',t)
	\nabla u(|\mathbf r-\mathbf r'|)$
	is the interaction-force density. The appearance of the two-body density
	$\psi^{(2)}$ is the point at which a closure is required.
	
	We employ the standard adiabatic approximation of DDFT: at each instant,
	the nonequilibrium interaction correlations are replaced by those of an
	equilibrium reference system constrained to reproduce the instantaneous
	one-body density $\psi$. In a general position--orientation
	description this equilibrium reference system obeys both a translational
	force-balance relation and a rotational torque-balance relation.
	The corresponding equilibrium force and torque
	balances can be written compactly as
	\begin{align}
		-D_t\nabla\psi
		-\mu\psi\nabla V_{\rm ext}
		+\mu\mathbf F_{\rm int}^{\rm ad}
		&=
		-\mu\psi\nabla
		\frac{\delta\mathcal F[\psi]}{\delta\psi},
		\nonumber\\
		-D_r\partial_\theta\psi
		+\mu_r\mathcal T_{\rm int}^{\rm ad}
		&=
		-\mu_r\psi\partial_\theta
		\frac{\delta\mathcal F[\psi]}{\delta\psi}.
		\label{eq:adiabatic_balances}
	\end{align}
	Here $\mathbf F_{\rm int}^{\rm ad}=
	-
	\int d\mathbf r'\,d\theta'\,
	\psi_{\rm ad}^{(2)}
	(\mathbf r,\theta,\mathbf r',\theta',t)
	\nabla
	u(\mathbf r,\theta,\mathbf r',\theta')$ and
	$\mathcal T_{\rm int}^{\rm ad}=
	-
	\int d\mathbf r'\,d\theta'\,
	\psi_{\rm ad}^{(2)}
	(\mathbf r,\theta,\mathbf r',\theta',t)
	\partial_\theta
	u(\mathbf r,\theta,\mathbf r',\theta')$ denote the conservative interaction-force
	and interaction-torque densities of the adiabatic reference system, and
	$\mu_r$ is the rotational mobility associated with the passive
	orientational dynamics.
	
	For the present model the external confinement and the Gaussian pair
	interaction depend only on particle positions. They therefore generate no
	conservative orientational torque:
	$\partial_\theta V_{\rm ext}=0$ and
	$\mathcal T_{\rm int}^{\rm ad}=0$. The passive angular contribution is
	consequently reduced to ordinary rotational diffusion,
	$-\mu_r\psi\,\partial_\theta(\delta\mathcal F/\delta\psi)
	=-D_r\partial_\theta\psi$. This should be distinguished from the
	self-alignment torque, which is intrinsically nonequilibrium and is not
	generated by the free-energy functional.
	
	The equilibrium Helmholtz free energy is decomposed into ideal, external,
	and excess contributions, which are given in Eqs.~\eqref{eq:free_energy_main_a} to \eqref{eq:free_energy_main}. 
	For the bounded
	Gaussian potential we employ the mean-field form
	$\mathcal F_{\rm exc}^{\rm MF}$, which yields the interaction potential
	$\Phi_{\rm int}(\mathbf r,t)
	=\int d\mathbf r'\,
	u(|\mathbf r-\mathbf r'|)\rho(\mathbf r',t)$ so that $
	\frac{\delta\mathcal F}{\delta\psi}
	=
	k_{\rm B}T
	\ln(\Lambda^2\psi)
	+
	V_{\rm ext}
	+
	\Phi_{\rm int}
	$
	.
	The adiabatic and mean-field approximations play distinct roles: the
	adiabatic approximation replaces the instantaneous nonequilibrium
	interaction force by that of an equilibrium reference state, whereas the
	mean-field approximation specifies the excess free-energy functional used
	for that reference system. 
	
	With $D_t=\mu k_{\rm B}T$, the resulting translational and angular
	currents take the compact form
	\begin{subequations}
	\begin{align}
		\mathbf J
		&=
		v_0\hat{\mathbf n}\psi
		-
		D_t\nabla\psi
		-
		\mu\psi\nabla
		\left(
		V_{\rm ext}+\Phi_{\rm int}
		\right),
		\\
		J_\theta
		&=
		\omega_{\rm align}\psi
		-
		D_r\partial_\theta\psi .
		\label{eq:closed_currents_methods}
	\end{align}
	\end{subequations}
	The nonequilibrium ingredient specific to the present system is retained
	through the velocity-based self-alignment rule. At the one-body level, the
	microscopic velocity entering the torque is represented by the local
	transport velocity,
	\begin{equation}
		\omega_{\rm align}(\mathbf r,\theta,t)
		=
		\lambda
		\left[
		\hat{\mathbf n}(\theta)
		\times
		\frac{\mathbf J_\rho(\mathbf r,t)}
		{\rho(\mathbf r,t)}
		\right]\cdot\hat{\mathbf e}_z,
		\label{eq:alignment_closure_methods}
	\end{equation}
	where $\mathbf J_\rho
	=
	\int_0^{2\pi}d\theta\,\mathbf J$. Thus the conservative forces are treated through an adiabatic
	density-functional closure, whereas self-alignment remains an explicitly
	active contribution. This distinction is central to the theory: the
	instantaneous density and orientational distribution determine the local
	particle current, the current reorients the particles through
	self-alignment, and the resulting change of orientation reconstructs the
	active current. The closed feedback between transport and orientation is
	the mechanism through which the DDFT captures the collective orbital
	dynamics observed in the particle simulations.

	\subsubsection*{Numerical solution of the DDFT}
	\label{methods:numerical_ddft}

	We solve the DDFT directly in the three-dimensional configuration space
	$(x,y,\theta)$ using a uniform Cartesian grid in position and a uniform
	angular grid on $\theta\in[0,2\pi)$. The continuity equation is discretized
	in conservative finite-volume form. The self-propulsion fluxes in the
	spatial directions and the self-alignment flux in orientation space are
	evaluated with an upwind discretization according to the sign of the
	corresponding advective velocity. Free-energy-driven fluxes are evaluated
	at cell faces using centred finite differences and arithmetic averages of
	the neighbouring densities.
	
	Periodic indexing is employed for the numerical grids in $x$, $y$, and
	$\theta$. The angular periodicity is physical, whereas the spatial domain is
	chosen sufficiently large that the harmonic confinement suppresses the
	density to numerically negligible values before the boundaries are reached.
	The spatial periodicity required by the discrete implementation therefore
	does not affect the confined dynamics.
	
	At each time step we first evaluate the orientationally integrated density
	\begin{equation}
		\rho(\mathbf r,t)
		=
		\int_0^{2\pi}
		d\theta\,
		\psi(\mathbf r,\theta,t).
		\label{eq:numerical_rho}
	\end{equation}
	The mean-field potential,
	$\Phi_{\rm int}=u*\rho$, is a two-dimensional spatial convolution and is
	computed efficiently using fast Fourier transforms. The functional
	derivative of $\mathcal F$ (Eqs.~\eqref{eq:free_energy_main_a} to \eqref{eq:free_energy_main}) is then evaluated over the
	full $(x,y,\theta)$ grid and used to construct the translational current.
	Integration over orientation gives $\mathbf J_\rho$, from which the local
	transport velocity entering the alignment term is obtained.
	
	Because both $\rho$ and $\mathbf J_\rho$ become exponentially small in the
	far tails of the confined density, direct evaluation of
	$\mathbf J_\rho/\rho$ can become numerically ill-conditioned there. We
	therefore regularize the denominator by a small positive density scale,
	\begin{equation}
		\mathbf u_{\rm num}
		=
		\frac{\mathbf J_\rho}
		{\rho+\rho_{\rm reg}},
		\qquad
		\rho_{\rm reg}\ll\rho_{\rm max},
		\label{eq:velocity_regularization}
	\end{equation}
	chosen sufficiently small that it modifies only regions with negligible
	particle density. The resulting velocity field is used in
	Eq.~\eqref{eq:ddft_rot_current} to calculate the angular advective flux.
	
	Time integration is performed explicitly. The time step is chosen below the
	stability limits associated with spatial self-propulsion, angular
	self-alignment, translational diffusion, and rotational diffusion. After
	each update a small positive density floor is imposed to avoid numerical
	underflow in exponentially dilute regions, followed by normalisation such
	that
	\begin{equation}
		\int d\mathbf r
		\int_0^{2\pi}d\theta\,
		\psi(\mathbf r,\theta,t)
		=
		N
		\label{eq:number_conservation}
	\end{equation}
	is maintained to numerical precision. Convergence was checked by reducing
	the time step and independently refining the spatial and angular grids.
	The reported centre-of-mass trajectories, dynamical-state assignments, and
	orbital loop numbers were unchanged under these refinements.
	
	For presentation, lengths are measured in units of the interaction range
	$\sigma$ and times in units of the orientational persistence time
	$\tau=D_r^{-1}$. The dimensionless self-alignment strength is therefore
	$B=\beta\sigma/\gamma_r=\lambda\sigma$, while the trap anisotropy is
	$\alpha=k_y/k_x$.
	
	The initial state is a spatially localized Gaussian density distribution
	centred away from the minimum of the trap,
	\begin{equation}
		\psi(\mathbf r,\theta,0)
		\!=\!
		\frac{\rho_0(\mathbf r)}{2\pi},
		\qquad
		\rho_0(\mathbf r)
		\!=\!
		A
		\exp\!\left[
		-\frac{|\mathbf r-\mathbf r_0|^2}
		{2w_0^2}
		\right] \!,
		\label{eq:initial_ddft}
	\end{equation}
	where $A$ is fixed by
	$\int d\mathbf r\,\rho_0(\mathbf r)=N$,
	$w_0$ determines the initial cloud width, and
	$\mathbf r_0$ specifies its displacement from the trap centre. The
	orientational factor $1/(2\pi)$ ensures vanishing initial polarization. The
	displaced cloud breaks the spatial symmetry of the preparation while
	leaving the direction of propulsion and the sense of any subsequently
	formed orbit dynamically undetermined.
	
	During the evolution we record the density, density current, orientational
	distribution, and centre-of-mass motion. The centre-of-mass velocity can be
	obtained either by differentiating its position or directly from the
	integrated current,
	\begin{equation}
		\mathbf V_{\rm CM}(t)
		=
		\frac{1}{N}
		\int d\mathbf r\,
		\mathbf J_\rho(\mathbf r,t).
		\label{eq:cm_velocity_methods}
	\end{equation}
	The resulting centre-of-mass time series provides the common input for the
	classification of the collective dynamical states and for the topological
	analysis of the orbital trajectories described below.


	\subsection*{Topological analysis of centre-of-mass orbits}
	\label{methods:topology_cm}
	The same topological analysis is applied to the BD and DDFT trajectories.
	For BD,
	$\mathbf R_{\rm CM}(t)=N^{-1}\sum_i\mathbf r_i(t)$, while for DDFT the
	centre of mass is obtained from the first moment of $\rho(\mathbf r,t)$.
	All subsequent steps therefore operate only on the two-dimensional time
	series $\mathbf R_{\rm CM}(t)
	=
	\bigl(x_{\rm CM}(t),y_{\rm CM}(t)\bigr)$.
	
	We first discard the initial $25\%$ of the recorded centre-of-mass trajectory.
	The orbital period is then estimated from the final $60\%$ of the remaining
	signal. After subtracting the corresponding late-time means,
	$\delta x_{\rm CM}
	=
	x_{\rm CM}-\langle x_{\rm CM}\rangle_{\rm lt},
	$ and
	$	\delta y_{\rm CM}
	=
	y_{\rm CM}-\langle y_{\rm CM}\rangle_{\rm lt}$, we determine the dominant positive
	Fourier frequencies
	\begin{equation}
		f_x^\star
		=
		\underset{f>0}{\operatorname{arg\,max}}
		|\widetilde{\delta x}_{\rm CM}(f)|,
		\qquad
		f_y^\star
		=
		\underset{f>0}{\operatorname{arg\,max}}
		|\widetilde{\delta y}_{\rm CM}(f)| ,
		\label{eq:dominant_cm_frequencies}
	\end{equation}
	using a Hann window to reduce spectral leakage. For the periodic loop states
	considered here, the Cartesian frequencies are commensurate harmonics of the
	same collective motion. We therefore take
	\begin{equation}
		f_0=\min(f_x^\star,f_y^\star),
		\qquad
		T_0=f_0^{-1},
		\label{eq:fundamental_period}
	\end{equation}
	using the non-vanishing component if only one Cartesian coordinate oscillates.
	The representative orbit is the final interval
	\begin{equation}
		\Gamma
		=
		\{
		\mathbf R_{\rm CM}(t):
		t_f-T_0\le t\le t_f
		\}.
		\label{eq:representative_cycle}
	\end{equation}
	Restricting the analysis to one period prevents repeated traversal of the same
	closed curve from trivially multiplying its winding number.
	
	For the numerical winding calculation, $\Gamma$ is represented by $N_\Gamma$ ordered
	points $\mathbf R_j=(x_{{\rm CM},j},y_{{\rm CM},j})$ and closed by joining the last point to the
	first. For a test point $\mathbf a$, define
	$\mathbf q_j=\mathbf R_j-\mathbf a$. The winding number is evaluated from the
	accumulated signed angular increment,
	\begin{equation}
		w(\mathbf a)
		=
		\frac{1}{2\pi}
		\sum_{j=0}^{N_\Gamma-1}
		\operatorname{atan2}
		\left[
		\mathbf q_j\times\mathbf q_{j+1},
		\mathbf q_j\cdot\mathbf q_{j+1}
		\right],
		\label{eq:discrete_winding}
	\end{equation}
	with $\mathbf R_{N_\Gamma}\equiv\mathbf R_0$. In two dimensions,
	$\mathbf q_j\times\mathbf q_{j+1}$ denotes the scalar out-of-plane component.
	This discrete form directly measures the accumulated angular rotation and
	avoids numerical differentiation of the trajectory.
	
	The winding field is evaluated on a $170\times170$ Cartesian grid extending
	by $18\%$ beyond the largest spatial span of the selected orbit. Numerical
	winding values are projected onto the nearest integer, and connected regions
	with identical nonzero winding are identified using eight-neighbour
	connectivity. Points on the trajectory itself are boundaries between winding
	domains and have no independent winding number.
	
	If the resulting connected domains are denoted by $\mathcal D_\ell$, the
	orbital loop number is
	\begin{equation}
		\mathcal Q_{\rm loop}
		=
		\sum_\ell |w_\ell|.
	\end{equation}
	Oppositely oriented lobes therefore contribute independently; for example,
	a figure-eight with $w=-1$ and $w=+1$ in its two lobes gives
	$\mathcal Q_{\rm loop}=2$. The identical transient removal, period
	extraction, winding calculation and domain analysis are used for BD and
	DDFT.
	\subsection*{Centre-of-mass probability-current reconstruction}
	\label{methods:flux_cm}
	The probability-current analysis uses the retained late-time CM trajectory
	after the same transient exclusion employed for the orbital-topology
	analysis. The sampled pairs $\{\mathbf R_{\rm CM}(t),\mathbf V_{\rm CM}(t)\}$
	are coarse-grained in the CM plane to estimate the probability density and
	probability current defined in Eq.~\eqref{eq:cm_probability_current}. To
	isolate the recirculating component, we represent the solenoidal current by
	a scalar stream function $\Psi_{\rm CM}$,
	\begin{equation}
		J^{\rm sol}_{{\rm CM},x}
		=
		\partial_{y_{\rm CM}}\Psi_{\rm CM},
		\qquad
		J^{\rm sol}_{{\rm CM},y}
		=
		-\partial_{x_{\rm CM}}\Psi_{\rm CM},
	\end{equation}
	which is obtained from
	\begin{equation}
		\nabla^2\Psi_{\rm CM}
		=
		-\bigl[\nabla\times\mathbf J_{\rm CM}\bigr]_z .
		\label{eq:cm_streamfunction_poisson}
	\end{equation}
	With this sign convention, positive and negative extrema correspond to
	counter-clockwise and clockwise circulation, respectively. Connected extrema
	of $\Psi_{\rm CM}$ define the coherent recirculating cells used to determine
	$N_{\rm cell}$.
	
	To compare the cellular structure between BD and DDFT independently of small
	differences in the overall extent and orientation of the CM distribution, we
	introduce principal coordinates. Let $\hat{\mathbf e}_s$ be the eigenvector of
	the late-time CM covariance matrix associated with its largest eigenvalue and
	$\hat{\mathbf e}_q$ the perpendicular direction. With
	$\Delta\mathbf R_{\rm CM}=\mathbf R_{\rm CM}-\langle\mathbf R_{\rm CM}\rangle_t$,
	we define
	\begin{equation}
		s=\hat{\mathbf e}_s\!\cdot\Delta\mathbf R_{\rm CM},
		\qquad
		q=\hat{\mathbf e}_q\!\cdot\Delta\mathbf R_{\rm CM}.
		\label{eq:principal_cm_coordinates}
	\end{equation}
	For states containing two or more cells, the longitudinal coordinate is
	normalised as
	\begin{equation}
		\hat s=\frac{s-s_0}{\ell_s},
		\qquad
		s_0=\frac{s_{\rm L}+s_{\rm R}}{2},
		\qquad
		\ell_s=\frac{s_{\rm R}-s_{\rm L}}{2},
		\label{eq:normalized_principal_coordinate}
	\end{equation}
	where $s_{\rm L}$ and $s_{\rm R}$ are the longitudinal positions of the two
	outermost cell centres. The outer cells are therefore mapped to
	$\hat s=-1$ and $+1$ while the relative positions of any internal cells are
	preserved. For the single-cell state, $s_0$ is taken at the maximum of the
	one-dimensional circulation profile and $\ell_s$ as its half width at half
	maximum.
	
	
	To compare the cellular structure of BD and DDFT quantitatively, we project $\Psi_{\rm CM}$ onto the principal CM direction and define the normalised longitudinal circulation profile
	\begin{equation}
		\overline{\Psi}_{\rm CM}(\hat s)
		=
		\frac{
			\displaystyle\int dq\,
			\left|\Psi_{\rm CM}(s_0+\ell_s\hat s,q)\right|
		}{
			\displaystyle\max_{\hat s}
			\int dq\,
			\left|\Psi_{\rm CM}(s_0+\ell_s\hat s,q)\right|
		},
		\label{eq:cm_circulation_profile}
	\end{equation}
where $s$, $q$, $s_0$ and $\ell_s$ are the principal coordinates and normalisation defined in Eqs.~\eqref{eq:principal_cm_coordinates}--\eqref{eq:normalized_principal_coordinate}. Each cell appears as a maximum of $\overline{\Psi}_{\rm CM}(\hat s)$, so that the number and relative placement of the cells can be compared between BD and DDFT independently of small differences in the extent of their CM distributions. The positive and negative contributions shown separately in Fig.~\ref{fig:cm_streamfunction}(e--h) are obtained by applying the same transverse integration to the corresponding sign sectors of $\Psi_{\rm CM}$. The same procedure is applied to BD and DDFT data.

	
	\subsection*{Internal-flow and rotational-activity analysis}
	\label{methods:internal_flow}
	The internal-flow analysis is performed in the instantaneous centre-of-mass
	frame defined in Eq.~\eqref{eq:internal_velocity}. For DDFT we use
	$\mathbf J_{\rho}(\mathbf r,t)=\int_0^{2\pi}d\theta\,\mathbf J(\mathbf r,\theta,t)$;
	for BD, the density and translational current are obtained by coarse-graining
	the particle positions and velocities onto the same spatial representation
	used for the field analysis. The vorticity and density-weighted enstrophy are
	then evaluated from Eqs.~\eqref{eq:internal_vorticity} and
	\eqref{eq:local_enstrophy_density}.
	
	For comparison between states with slightly different spatial extent, the
	longitudinal coordinate is normalised by the density distribution itself. We
	define the longitudinal line density
	\begin{equation}
		\rho_{\parallel}(x)=\int dy\,\rho(x,y),
		\label{eq:longitudinal_line_density}
	\end{equation}
	with centroid and root-mean-square width
	\begin{equation}
		x_{\rm c}
		=
		\frac{\int dx\,x\rho_{\parallel}(x)}
		{\int dx\,\rho_{\parallel}(x)},
		\qquad
		R_x^2
		=
		\frac{\int dx\,(x-x_{\rm c})^2\rho_{\parallel}(x)}
		{\int dx\,\rho_{\parallel}(x)}.
		\label{eq:longitudinal_extent}
	\end{equation}
	The normalised coordinate $\hat{x}=(x-x_{\rm c})/R_x$ therefore measures
	longitudinal position relative to the centre of the cloud in units of its own
	root-mean-square extent. The projected rotational-activity profile is then
	constructed from Eq.~\eqref{eq:longitudinal_enstrophy}. This normalisation is
	applied identically to BD and DDFT, allowing the spatial arrangement of
	vortical activity to be compared independently of small differences in the
	absolute cloud width.
	
	For the $\alpha=5$, $B=1$ BD data shown in
	Fig.~\ref{fig:vortex_enstrophy}, the longitudinal profile is averaged over
	realizations that reach the two-cell vortex-like state; the shaded band shows
	the standard error of the mean. The $\alpha=1$, $B=0.5$ BD field maps and
	profile are shown for a representative realization, as indicated in the figure
	caption.
	
	\subsection*{Brownian-dynamics simulations}
	\label{methods:bd_simulations}
	We complement the continuum theory by particle-resolved Brownian-dynamics
	(BD) simulations of the same interacting self-aligning active fluid. To
	retain a well-defined instantaneous particle velocity, which enters
	explicitly in the self-alignment mechanism, we employ a weakly inertial
	Langevin representation and work in the small-inertia regime. The
	translational velocity $\mathbf v_i=\dot{\mathbf r}_i$ and angular velocity
	$\omega_i=\dot{\theta}_i$ of particle $i$ obey
	\begin{subequations}
	\begin{align}
		m\dot{\mathbf v}_i
		&=
		-\gamma\mathbf v_i
		+\gamma v_0\hat{\mathbf n}_i
		+\mathbf F_i
		+\gamma\sqrt{2D_t}\,
		\boldsymbol\xi_i(t),
		\label{eq:bd_translation}
		\\
		I\dot{\omega}_i
		&=
		-\gamma_r\omega_i
		+
		\beta
		\left(
		\hat{\mathbf n}_i\times\mathbf v_i
		\right)\cdot\hat{\mathbf e}_z
		+
		\gamma_r\sqrt{2D_r}\,
		\eta_i(t),
		\label{eq:bd_rotation}
	\end{align}
	\end{subequations}
	together with
	$\dot{\mathbf r}_i=\mathbf v_i$ and
	$\dot{\theta}_i=\omega_i$.
	Here $m$ and $I$ are the particle mass and moment of inertia,
	$\gamma$ and $\gamma_r$ are the translational and rotational damping
	coefficients, and
	$\hat{\mathbf n}_i=(\cos\theta_i,\sin\theta_i)$ is the propulsion
	orientation. The active force $\gamma v_0\hat{\mathbf n}_i$ produces the
	free-particle propulsion speed $v_0$, while
	$\mathbf F_i=-\nabla_iU(\mathbf r^N)$ contains the Gaussian pair forces
	and anisotropic harmonic confinement defined in the main text.
	The parameter $\beta>0$ controls the self-aligning torque and is related
	to the alignment coefficient used in the continuum theory through
	$\lambda=\beta/\gamma_r$, so that the dimensionless alignment strength is
	$B=\beta\sigma/\gamma_r$.
	
	The translational and rotational noises,
	$\boldsymbol\xi_i=(\xi_{ix},\xi_{iy})$ and $\eta_i$, are independent
	Gaussian white noises with zero mean and correlations
	\begin{subequations}
	\begin{align}
		\left\langle
		\xi_{i\alpha}(t)\xi_{j\beta}(t')
		\right\rangle
		&=
		\delta_{ij}\delta_{\alpha\beta}\delta(t-t'),
		\\
		\left\langle
		\eta_i(t)\eta_j(t')
		\right\rangle
		&=
		\delta_{ij}\delta(t-t').
		\label{eq:bd_noise}
	\end{align}
	\end{subequations}
	The coefficients $D_t$ and $D_r$ set the translational and rotational
	noise strengths, respectively. For the translational fluctuations we use
	the Einstein relation $D_t=k_{\rm B}T/\gamma$. The rotational diffusivity
	$D_r$ is treated directly as the orientational-noise strength and fixes the
	persistence time
	$\tau=D_r^{-1}$.
	
	The finite values of $m$ and $I$ are introduced only to resolve the
	instantaneous velocity and angular velocity. Their dynamical relevance is
	controlled by the translational and rotational relaxation times
	$\tau_m=m/\gamma$ and $\tau_I=I/\gamma_r$. Relative to the persistence
	time, these define the reduced inertial parameters
	\begin{equation}
		M=\frac{mD_r}{\gamma},
		\qquad
		\mathcal I=\frac{ID_r}{\gamma_r}.
		\label{eq:reduced_inertia}
	\end{equation}
	Throughout the simulations we use
	$M=\mathcal I=10^{-3}$, such that
	$\tau_m,\tau_I\ll\tau$. The particle motion is therefore in the
	overdamped regime on the persistence and collective time scales of
	interest, while $\mathbf v_i$ remains an explicitly resolved dynamical
	quantity. This is particularly useful for the present model because the
	self-alignment torque responds directly to the instantaneous mismatch
	between $\hat{\mathbf n}_i$ and $\mathbf v_i$.
	
	Lengths and times are reported in units of $\sigma$ and
	$\tau=D_r^{-1}$, respectively. Besides the confinement anisotropy
	$\alpha=k_y/k_x$ and the self-alignment strength
	$B=\beta\sigma/\gamma_r$, the activity is characterised by the
	P\'eclet number
	${\rm Pe}=v_0/(D_r\sigma)$. Unless stated otherwise, we use
	\[
	N=400,\qquad
	\sigma=1,\qquad
	D_r=1,\qquad
	\gamma=\gamma_r=1,
	\]
	together with
	$v_0=50$, $D_t=0.1$, $\epsilon=500$, and $k_x=1$,
	corresponding to ${\rm Pe}=50$. The mass and moment of inertia are
	$m=I=10^{-3}$, while $k_y$ and $\beta$ are varied to control
	$\alpha$ and $B$, respectively. The simulations are performed in a square
	computational domain of side $60\sigma$. The harmonic confinement keeps
	the particle distribution well inside this domain, while periodic boundary
	conditions and the minimum-image convention are used for the numerical
	evaluation of particle separations.
	
	The stochastic equations are advanced with a first-order stochastic Euler
	scheme. At each time step the pair and confinement forces are first
	evaluated from the instantaneous particle configuration. The translational
	velocities are then updated from Eq.~\eqref{eq:bd_translation}, followed
	by the particle positions; analogously, the angular velocities are updated
	from Eq.~\eqref{eq:bd_rotation} before advancing the orientational angles.
	Independent Gaussian random numbers are generated for each particle,
	Cartesian component, and time step. We use
	$\Delta t=10^{-4}\tau$, which resolves the inertial relaxation times,
	and evolve the production trajectories up to $T=20\tau$.
	
	The initial particle positions are sampled from the same displaced Gaussian
	cloud used in the DDFT calculations, with centre
	$\mathbf r_0=(5\sigma,-5\sigma)$ and width $w_0=10\sigma$.
	The initial translational velocities are chosen along the instantaneous
	propulsion directions,
	$\mathbf v_i(0)=v_0\hat{\mathbf n}_i(0)$, while
	$\omega_i(0)=0$. The initial particle orientations are drawn uniformly at random over $\theta\in[0,2\pi)$, corresponding to an isotropic orientational distribution and matching the $1/(2\pi)$ factor in Eq.~\eqref{eq:initial_ddft}, which is consistent with the continuum initial condition.
	
	For comparison with the continuum fields, particle configurations are
	coarse-grained on a $256\times256$ Cartesian grid, and orientational
	distributions are resolved using $128$ angular bins. The particle density,
	polarization density, and translational current are constructed directly
	from the instantaneous particle positions, orientations, and velocities.
	The centre-of-mass position and velocity are evaluated as
	$\mathbf R_{\rm CM}=N^{-1}\sum_i\mathbf r_i$ and
	$\mathbf V_{\rm CM}=N^{-1}\sum_i\mathbf v_i$, respectively. Ensemble
	quantities are obtained from ten statistically independent realizations.
	The resulting particle-resolved trajectories and fields are analysed using
	the same dynamical-state and orbital-topology criteria as the DDFT data.

	\section*{DATA AVAILABILITY}
	The data and codes supporting the findings of this study are publicly available on Zenodo at \href{https://doi.org/10.5281/zenodo.24003960}{https://doi.org/10.5281/zenodo.24003960}.
	

	\section*{VIDEO DESCRIPTION}

\noindent
Supplementary Videos~1--4 illustrate the four representative dynamical states
marked by stars in Fig.~\ref{figure02}. Each movie shows the same
state through complementary particle-resolved (labeled as \textbf{a}) and density-field (labeled as \textbf{b}) representations.

\vskip5pt
\noindent
\textbf{Supplementary Video 1:} Arrested state. The active cluster remains
localized around a nearly stationary centre of mass, with no sustained
large-scale orbital motion.

\vskip5pt
\noindent
\textbf{Supplementary Video 2:} Vortex-like state. Global translation is
strongly suppressed while the active material sustains pronounced internal
circulation around an almost stationary centre of mass.
\vskip5pt
\noindent
\textbf{Supplementary Video 3:} Stable-loop state. The cluster develops
persistent collective orbital motion and converges to a reproducible closed
centre-of-mass trajectory.

\vskip5pt
\noindent
\textbf{Supplementary Video 4:} Oscillatory state. The cluster undergoes
persistent back-and-forth collective motion, producing a recurrent but
non-orbiting centre-of-mass trajectory.

	%

	\section*{References}
	
	\vspace*{10pt}
	
	\begin{acknowledgements}
		IA acknowledges funding support by the Deutsche Forschungsgemeinschaft (DFG) under project number 556762905 — AB 1083/1-1.  LC acknowledges funding from the Italian Ministero dell’Università e della Ricerca under the programme PRIN 2022 ("re-ranking of the final lists"), number 2022KWTEB7, cup B53C24006470006.
		HL acknowledges support by the Deutsche Forschungsgemeinschaft (DFG) through the SPP 2265, under grant
		number LO418/25. \\

	\end{acknowledgements}
	
	\subsection*{Author contributions}
	IA, MM and HL conceived and  conceptualised the research. IA did the analytical calculations and their analysis. MM and IA performed numerical simulations and analyzed the results. IA, LC, and HL wrote the first draft of the paper. All authors discussed and interpreted the results and wrote the final version of the manuscript.  
	
	\subsection*{Competing interests}
	The authors declare no competing interests. \\
	
	\subsection*{Additional information}
	\textbf{Supplementary information} The online version contains
	supplementary material available at (the address will be provided by the journal) \\
	
	\textbf{Correspondence} and requests for materials should be addressed to
	Iman Abdoli.
	
\end{document}